\documentclass[twocolumn,showpacs,amsfonts,aps,prc,nofootinbib,floatfix,superscriptaddress]{revtex4}

\usepackage{bm}
\usepackage{graphicx}
\usepackage{amsmath}
\usepackage{epstopdf}
\usepackage{xcolor}
\usepackage[normalem]{ulem}

\def\l{\left}
\def\r{\right}

\def\BIGGL{\scalebox{1.25}{\Bigg[}}
\def\BIGGR{\scalebox{1.25}{\Bigg]}}

\def\ev{{\mathrm{ev}}}
\def\mcO{{\mathcal{O}}}
\def\mcP{{\mathcal{P}}}
\def\mco{o}
\def\Mmax{M_{\mathrm{max}}}

\def\nev{N_{\mathrm{ev}}}
\newcommand{\azavg}[1]{\left< #1 \right>_{\Phi_K}} 
\newcommand{\evavg}[1]{\left< #1 \right>_{\ev}} 

\newcommand{\e}{\mbox{e}}
\newcommand{\erf}{\mbox{erf}}

\newcommand{\var}[1]{\mbox{Var} \left[ #1 \right]}
\newcommand{\cov}[2]{\mbox{Cov} \left( #1, #2 \right)}
\newcommand{\avg}[1]{\left< #1 \right>} 

\begin{document}


\title{Probing the properties of event-by-event distributions in Hanbury-Brown--Twiss radii}

\author{Christopher Plumberg}
\author{Ulrich Heinz}
\affiliation{Department of Physics, The Ohio State University,
  Columbus, OH 43210-1117, USA}

\begin{abstract}
Hanbury-Brown--Twiss interferometry is a technique which yields effective widths (i.e., ``HBT radii") of homogeneity regions in the fireballs produced in heavy ion collisions.  Because the initial conditions of these collisions are stochastically fluctuating, the measured HBT radii also exhibit variation on an event-by-event basis.  However, HBT measurements have, to date, been performed only on an ensemble-averaged basis, due to inherent limitations of finite particle statistics.  In this paper, we show that experimental measurements to date are best characterized theoretically as weighted averages of the event-by-event HBT radii, and we propose a new method for extracting experimentally both the arithmetic mean and the variance of the event-by-event distribution of HBT radii.  We demonstrate the extraction of the mean and variance of this distribution for a particular ensemble of numerically generated events, and offer some ideas to extend and generalize the method to enable measurement of higher moments of the HBT distribution as well.
\end{abstract}


\date{\today}

\maketitle

\section{Introduction}
\label{sec1}

HBT interferometry relies on two-particle momentum correlations to extract information about the spatiotemporal structure of the emitting source in heavy-ion collisions.  The technique depends on the detection of pairs of identical particles (e.g., pions or kaons), whose quantum statistical correlations convey important information about the mean relative separation between the points at which the particles were emitted during the freeze-out process.  Ideally, one would be able to do this on an event-by-event basis: if any given collision yielded a sufficiently large number of the desired particles in the final state, this would allow a measurement of the HBT radii, the effective widths of the homogeneity regions in the fireball, event by event.  Unfortunately, after the total particle multiplicity (on the order of a few 1000 per event) is binned according to particle species, $p_T$, and emission angle, not enough pairs remain for a statistically meaningful, fully three dimensional analysis of the correlation function.

Consequently, experimentalists typically combine large numbers ($\gtrsim 10^6$) of events in order to boost the pair statistics, thereby increasing the precision of the resulting HBT measurements.  The collection of events is referred to as the \textit{ensemble}, and the two-particle correlation function (from which the HBT radii are experimentally extracted) thus contains a non-trivial combination of the correlation functions of all of the events in the ensemble.  An apples-to-apples comparison with theoretical models therefore requires, at least in principle, a corresponding ensemble averaging on the theoretical side.

The process of ensemble averaging has historically been accounted for at the level of the initial state of the fireball.  In its crudest form, the ensemble of fluctuating events is replaced by a single averaged event whose final state is computed by hydrodynamically evolving a single averaged initial profile.  With the recent availability of resources to evolve large numbers of collisions with fluctuating initial conditions event by event, the ensemble averaging procedure has been shifted to the emission function.  By performing an ensemble average directly over emission functions constructed from the freeze-out surfaces of each event in the ensemble, the two particle correlation function can be related to the Fourier transform of the ensemble-averaged emission function.  However, since the experimental correlation function is constructed after the final state particles have been emitted from the freeze-out surfaces of each respective event, it is more accurate to perform the ensemble-averaging procedure at the level of the correlation function itself.  This induces corrections to the HBT radii extracted from the correlation function which is constructed from the ensemble-averaged emission function, and these corrections are sensitive to event-by-event fluctuations encoded in the structure of the freeze-out surface.  Only this last procedure invokes a \textit{distribution} of correlation functions and thus a distribution of HBT radii which can be characterized by a mean, a variance, and possible higher moments.  In this paper we will analyze which moment of this distribution is represented by the experimentally measured HBT radii, and what additional measurements could be made to access other moments of the HBT radii distribution.

Numerical studies such as \cite{Bozek:2014hwa} have shown that the mean HBT radii extracted from a fluctuating set of correlation functions are almost indistinguishable from the radii characterizing the single correlation function obtained from an ensemble-averaged emission function.  On the other hand, recent event-by-event simulations \cite{Plumberg:2015eia} indicate that the HBT radii extracted from individual events\footnote{
	This is possible in theory since the correlation function does not need to be sampled
	with a finite number of particles but can be calculated with infinite precision.
	}
may fluctuate with a typical range of 10-15\% (in the squared HBT radii) for central collisions and that, if these fluctuations are both present in actual heavy-ion collisions and experimentally accessible, their scale could provide valuable sensitivity, e.g., to different functional forms of the $T$-dependence in $\eta/s$.  In this paper, we propose a method for extracting the scale of these fluctuations experimentally.  A more thorough exploration of our method's theoretical implications is deferred to another work.

The layout of the remainder of this paper is as follows.  In section \ref{sec2}, we introduce the formalism underlying both event-by-event and ensemble-averaged HBT analyses which retain the correlation function's intrinsic dependence on the pair emission angle, and distinguish several different, commonly used methods for theoretically computing the ensemble-averaged HBT radii which are measured experimentally.  We argue that each of these methods traces qualitatively, but not with quantitative precision, the arithmetic mean of the event-by-event distribution of the HBT radii.  In section \ref{sec3}, we discuss analogous results for HBT analyses in which the dependence on the pair emission angle is averaged over, and show how this simplification affects the differences between the various methods of ensemble averaging.  The best theoretical representation of the experimentally employed ensemble averaging process identifies the measured HBT radii as weighted averages of the event-by-event radii.  We discuss these weights.  With this in mind, we show in section \ref{sec5} how to estimate the first moment (i.e., the mean) of an event-by-event distribution in terms of linear combinations of such weighted averages.  Although we focus in this paper on the HBT radii, the methods we introduce are quite general and applicable to any event-by-event observables.  In sections \ref{sec6} and \ref{sec7}, we show how to access higher moments of the event-by-event distribution by performing repeated averages over sub-ensembles and measuring their fluctuations; in section \ref{sec6}, we concentrate on estimating the variance, while section \ref{sec7} addresses higher moments.  Finally, in section \ref{sec8}, we present a proof-of-principle demonstration of our method and discuss some subtleties relevant for its experimental implementation.  A detailed derivation of our method (for estimating the variance) is provided in Appendix \ref{App:AppendixA}.

\section{Azimuthally sensitive HBT interferometry for fluctuating sources}
\label{sec2}

In this section we discuss HBT interferometry that is fully differential in the pair momentum $\vec{K}$, in particular its azimuthal angle $\Phi_K$ around the beam direction.  This is known as ``azimuthally sensitive HBT interferometry" \cite{Wiedemann:1997cr,Lisa:2000ip}.  In Sec. \ref{sec3} we will modify the treatment for azimuthally averaged (i.e., $\Phi_K$-integrated) measurements.

\subsection{Azimuthally sensitive interferometry for a single fluctuating event}
\label{sec2a}

HBT interferometry is founded on the concept of the two-particle correlation function, defined for a single event by
\begin{equation}
C(\vec{p}_1, \vec{p}_2) \equiv \frac{ E_{p_1} E_{p_2} \frac{d^6 N}{d^3 p_1 d^3 p_2}}{ \l( E_{p_1} \frac{d^3 N}{d^3 p_1}\r) \l(  E_{p_2} \frac{d^3 N}{d^3 p_2} \r) }. \label{corrfuncSEdefn}
\end{equation} 
Here, $\vec{p}_1$ and $\vec{p}_2$ represent the 3-momenta of identical particles (e.g., pions) which have been emitted from the fireball.  The correlation function \eqref{corrfuncSEdefn} may be interpreted as the probability of simultaneously measuring two particles with momenta $\vec{p}_1$ and $\vec{p}_2$ in a single event, divided by the probability of measuring the same two particles (with the same momenta) \textit{independently} in two separate but identical events.  Correlations among the particles in the emitted pair manifest themselves as deviations of $C(\vec{p}_1, \vec{p}_2)$ from unity.  The connection of $C(\vec{p}_1, \vec{p}_2)$ with the size of the effective emission region ("homogeneity region") from which the pairs are emitted is provided by the following connection \cite{Heinz:1999rw} with the single-particle Wigner density (or "emission function") of the fireball, $S(x,K)$:\footnote{The '+' sign corresponds to using bosons to construct the correlator, whereas the '-' corresponds to using fermions.}
\begin{equation}
C(\vec{q}, \vec{K}) \approx  1 \pm \l| \frac{ \int d^4 x \, S(x,K) \e^{iq \cdot x} }{\int d^4 x \, S(x,K)} \r|^2, \label{corrfunc_vs_S_defn}
\end{equation} 
where we have introduced the notation $q^{\mu} \equiv p^{\mu}_1 - p^{\mu}_2$, $K^{\mu} \equiv (p^{\mu}_1 + p^{\mu}_2)/2$.  Eq.~\eqref{corrfunc_vs_S_defn} holds in the absence of final state interactions between the emitted particles and for ``chaotic" sources that emit the two particles independently from each other.  The approximation indicated by the $\approx$ sign refers to the replacement of $\vec{p}_1, \vec{p}_2$ by $\vec{K}$ in the denominator (the so-called ``smoothness approximation" \cite{Heinz:1999rw}).  If the pairs of identical particles used in the construction of the numerator and denominator of Eq.~\eqref{corrfuncSEdefn} are bosons (as we consider in this paper), the correlation function itself experiences an enhancement near $\vec{q}=0$; this enhancement is usually described by a functional form which is Gaussian in the components of the relative momentum $\vec{q}$:
\begin{equation}
C_{fit}(\vec{q}, \vec{K}) \equiv 1 + \lambda(\vec{K}) \exp \l(-\sum_{i,j = o, s, l} R^2_{ij}(\vec{K}) q_i q_j \r). \label{corrfunc_functional_defn}
\end{equation}
This Gaussian parametrization is exact for emission functions with a Gaussian spatial structure and is usually adequate for non-Gaussian sources whose deviations from Gaussian structure are generated by additional length scales characterizing the source that are very different from the source radii.  Here, $\lambda(\vec{K})$ (the ``intercept parameter") encodes information about long-lived resonances which decay well outside the reaction zone of the fireball, and is used to account for the  resulting empirical reduction in the peak value of $C$ at $\vec{q}=0$ \cite{Heinz:1999rw,Lisa:2005dd}.  We neglect the contributions from resonances and set $\lambda(\vec{K})=1$.  The sum in the exponent ranges over the coordinates of the widely used $osl$-system, where $l$ (the ``longitudinal" direction) coincides with the beam direction, $o$ (the ``outward" direction) points in the same direction as $\vec{K}_T$, the average pair momentum projected onto the transverse plane, and $s$ (the ``sideward" direction) points perpendicular to both of these.  In terms of these coordinates, Eq.~\eqref{corrfunc_functional_defn} defines the \textit{HBT radius parameters} $R^2_{ij}(\vec{K})$, whose diagonal components may be interpreted as the squares of the effective widths of the emission regions within the fireball responsible for producing particle pairs with average momentum $\vec{K}$.

One prescription for computing the HBT radii from theoretical (hydrodynamical) models on an event-by-event basis relies on the Cooper-Frye formula \cite{Cooper:1974mv} to define the eventwise emission function, and then uses \eqref{corrfunc_vs_S_defn} to define the corresponding correlation function.  This correlation function can then be fit using \eqref{corrfunc_functional_defn} to obtain the $R^2_{ij}$ for that event as fit parameters.  Explicitly, in the Cooper-Frye algorithm the emission function is defined as follows:
\begin{eqnarray}
S(x,p) &=& \frac{1}{(2\pi)^3} \int_{\Sigma} p \cdot d^3 \sigma(y)\, \delta^4 (x{-}y)\, f(y,p)\,, \label{cooper_frye_defn1}
\\
f(x,p) 	&=& f_0 \l(x,p\r) + \delta f \l(x,p\r)\nonumber\\
		&=& \frac{1}{e^{(p \cdot u{-} \mu)/T}{-}1} + \frac{p^{\mu} p^{\nu} \pi_{\mu\nu} }{2 T^2 (e{+}{\cal P})} f_0 (1{+}f_0). 
\label{cooper_frye_defn2}
\end{eqnarray}
Here, $\delta f$ is the first-order viscous correction to the local equilibrium distribution function $f_0$ \cite{Teaney:2003kp,Dusling:2009df}, and we assume a quadratic dependence on $p$. $\pi_{\mu\nu}(x)$ is the viscous pressure tensor, $u^{\mu}(x)$ is the flow velocity profile along the freeze-out surface, and $\mu$, $T_{\mathrm{dec}}$, $e$, and $\cal P$ are the chemical potential, decoupling temperature, energy density, and pressure, respectively, which, for the hydrodynamical simulations used in this work, will all be taken as constant along the freeze-out surface by construction.  $\Sigma$ represents the freeze-out hypersurface over which the integration is performed, and $d^3\sigma_{\mu}(x)$ is the outward pointing normal vector at the point $x$ on this surface.

This prescription for generating $R^2_{ij}$ from theoretical models for comparison with experimental data is computationally intensive: to construct the correlation function for a given event according to \eqref{corrfunc_vs_S_defn} requires multidimensional integrations over the freeze-out surface of the event in question (where $S(x,K)$ itself, in general, requires a similar integration if computed according the Cooper-Frye prescription \eqref{cooper_frye_defn1} and \eqref{cooper_frye_defn2}); these integrations, moreover, must be performed for a sufficiently large number of points in $\vec{q}$ and $\vec{K}$ that the fit to \eqref{corrfunc_functional_defn} can be carried out with acceptable accuracy and still yield useful results for comparison with experiment.  To do all of this for a large ensemble of events consequently places stringent demands on available computational resources.

Much of this numerical expense can be avoided by adopting the following often-used approximation.  By assuming that the emission function (and, consequently, the corresponding correlation function) for each event can be described exactly as a Gaussian in $x$ (corresp., $q$), the fit of \eqref{corrfunc_vs_S_defn} to \eqref{corrfunc_functional_defn} becomes an identity, and the $R^2_{ij}$ may be read off directly in terms of integrals over the emission functions of the ensemble of events \cite{Heinz:1999rw,Lisa:2005dd}:
\begin{equation}
R^2_{ij}(\vec{K}) = \avg{(\tilde{x}_i - \beta_i \tilde{t})(\tilde{x}_j - \beta_j \tilde{t})}_S,  \label{svHBT_defn}
\end{equation}
where $\tilde{x}_{\mu} = x_{\mu} - \avg{x_{\mu}}_S$, $\vec{\beta} = \vec{K}/E_K$, and
\begin{equation}
\avg{f(x)}_S \equiv \frac{\int d^4 x\, f(x)\, S(x,K)}{\int d^4 x\, S(x,K)}  
\label{source_integral}
\end{equation}  
Although still computationally intensive, the project of performing event-by-event HBT analyses has been cast into a much more tractable form by the use of \eqref{svHBT_defn} and \eqref{source_integral}, which drastically reduce the number of required calculations for computing the $R^2_{ij}$ from theoretical models.  Although for realistic (i.e., hydrodynamic) model sources there exist well-documented discrepancies between the ``Gaussian fit" method and ``source variances" method, these discrepancies will not affect the qualitative results that we discuss in this paper, which will be based on Eqs.~\eqref{svHBT_defn} and \eqref{source_integral}.

\subsection{Azimuthally sensitive interferometry for ensembles of fluctuating events}
\label{sec2b}

While theoretically well-defined on an event-by-event basis, single-event HBT interferometry is in general not practically possible, as explained in the introduction.  For this reason, experimentalists typically modify the definition of the correlation function to include in the numerator and denominator of \eqref{corrfuncSEdefn} pairs of pions from multiple events; the collection of all events combined in this way is known as the \textit{ensemble}, and the corresponding definition of the correlation function is
\begin{equation}
C_{\mathrm{avg}}(\vec{p}_1, \vec{p}_2) \equiv \frac{\l< E_{p_1} E_{p_2} \frac{d^6 N}{d^3 p_1 d^3 p_2} \r>_{\ev}}{ \l<E_{p_1} \frac{d^3 N}{d^3 p_1}\r>_{\ev} \l< E_{p_2} \frac{d^3 N}{d^3 p_2} \r>_{\ev} }, \label{corrfuncENSAVG0defn}
\end{equation} 
where the $\avg{\ldots}_{\ev}$ notation is shorthand for
\begin{equation}
\avg{X}_{\ev} \equiv \frac{1}{N_{\ev}} \sum^{N_{\ev}}_{i=1} X_i,  \label{DEA_obs_defn}
\end{equation}
i.e., an arithmetic average of the quantity $X$ over all events in the ensemble.  In the construction of the correlator according to Eq.~\eqref{corrfuncENSAVG0defn}, it is important to align the events according to some direction defined by the individual event, for example, the $n$th order flow angle $\Psi_n$ of charged hadrons.  This will result in an azimuthal dependence of the HBT radii relative to that flow angle $\Psi_n$.  In Appendix \ref{App:AppendixB}, we show that the radii extracted from a Gaussian fit of \eqref{corrfuncENSAVG0defn}, which we label by $R^2_{\avg{ij}}$, can be related to the event-by-event HBT radii by
\begin{equation}
R^2_{\avg{ij}}(\vec{K}) =  \frac{\avg{N^2(\vec{K}) R^2_{ij}(\vec{K})}_{\ev}}{\avg{N^2(\vec{K})}_{\ev}}, \label{true_R2ij_from_corrfunc}
\end{equation} 
where $N(\vec{K}) \equiv E_K (d^3 N/d^3 K)$ is the Lorentz-invariant yield of the particles of interest (in our case, pions) with momentum $\vec{K}$.  Theoretically, there are many different ways to generalize \eqref{corrfuncSEdefn} to ensembles containing multiple events.  In the rest of this subsection, we discuss several of these alternatives.

\subsubsection{Ensemble-averaged initial conditions -- single-shot hydrodynamics}
\label{sec2b1}

One method for extracting from hydrodynamic codes a correlation function containing averaged information from multiple events involves ensemble-averaging (possibly after proper alignment if the events are deformed) the initial entropy (or energy) densities of all events in the transverse plane, and using the resulting averaged density profile as a set of initial conditions for hydrodynamics.  At the end of the hydrodynamic evolution, Eq.~\eqref{corrfunc_functional_defn} relates the emission function constructed on the freeze-out surface for this averaged density profile to a correlation function (by \eqref{corrfunc_vs_S_defn}) which is effectively insensitive to the existence of event-by-event fluctuations.  We refer to this method as "single-shot hydrodynamics", and we refer to the emission function (resp., correlation function) so constructed as $S_{\mathrm{ssh}}$ (resp., $C_{\mathrm{ssh}}$), and denote the HBT radii extracted with this method by $R^2_{\bar{ij}}$.  Explicitly, we may write (using the shortcut discussed above)
\begin{equation}
R^2_{\bar{ij}}(\vec{K}) = \avg{(\tilde{x}_i - \beta_i \tilde{t})(\tilde{x}_j - \beta_j \tilde{t})}_{\mathrm{ssh}}, \label{R2ij_from_SSH}
\end{equation} 
where $\avg{\ldots}_{\mathrm{ssh}}$ is defined with $S_{\mathrm{ssh}}$ as the weight function.

\subsubsection{Ensemble-averaged emission function}
\label{sec2b2}

Another common way of computing an ensemble-averaged correlation function consists of averaging the emission functions after the event-by-event hydrodynamic evolution of many events with fluctuating initial conditions.  In particular, we define $\bar{S}(x,K) \equiv \avg{S(x,K)}_{\ev}$ and, in analogy with \eqref{corrfunc_vs_S_defn}, introduce
\begin{equation}
\bar{C}(\vec{q}, \vec{K}) \equiv  1 + \l| \frac{ \int d^4 x \, \bar{S}(x,K) \e^{iq \cdot x} }{\int d^4 x \, \bar{S}(x,K)} \r|^2 \label{Cbar_corrfunc_def}
\end{equation}  
These definitions of $\bar{S}$ and $\bar{C}$ (and the corresponding radii, which we denote $\bar{R}^2_{ij}$) have an advantage over the corresponding ``single-shot hydrodynamics" definitions that we discussed above, in that the quantum fluctuations in the initial state are allowed to modify the hydrodynamic evolution event by event before being averaged over at the end.  Since the hydrodynamic evolution is nonlinear, $\bar{S}(x,K) \neq S_{\mathrm{ssh}}(x,K)$.  In this work, we will refer to this method as the "average emission function" method, and define
\begin{equation}
\bar{R}^2_{ij}(\vec{K}) = \avg{(\tilde{x}_i{-}\beta_i \tilde{t})(\tilde{x}_j{-}\beta_j \tilde{t})}_{\bar{S}}. \label{R2ij_from_Sbar}
\end{equation}
%

\subsubsection{Ensemble-averaged correlation function}
\label{sec2b3}

Of the available theoretical techniques for treating the ensemble-averaging process, the ``single-shot hydrodynamics" and ``average emission function" methods have historically enjoyed the greatest popularity.  However, as Eq.~\eqref{corrfuncENSAVG0defn} shows, the way to correctly reproduce the experimental process of performing the ensemble average is by first Fourier transforming the emission function and then averaging over events:
\begin{eqnarray}
C_{\mathrm{avg}}(\vec{p}_1, \vec{p}_2) &\equiv & \avg{C}_{\ev}(\vec{q}, \vec{K}) \nonumber\\
	&\equiv & 1 + \frac{ \l< \l| \int d^4 x \, S(x,K) \e^{iq \cdot x} \r|^2 \r>_{\ev}}{\l| \int d^4 x \, \bar{S}(x,K) \r|^2} \label{Cavg_corrfunc_def}\\
	&=& \bar{C}(\vec{q}, \vec{K}){+}\frac{ \l< \l| \int d^4 x \, \delta S(x,K) \e^{iq \cdot x} \r|^2 \r>_{\ev}}{\l| \int d^4 x \, \bar{S}(x,K) \r|^2}, \nonumber
\end{eqnarray}
where $\delta S(x,K) \equiv S(x,K) - \bar{S}(x,K)$.  Using the shortcut \eqref{svHBT_defn} we can then write the corresponding HBT radii $R^2_{\avg{ij}}$ extracted from a Gaussian fit of $\avg{C}_{\ev}(\vec{q}, \vec{K})$ as follows (see Eq.~\eqref{true_R2ij_from_corrfunc}):
\begin{equation}
	R^2_{\avg{ij}}(\vec{K})
		= \frac{1}{\avg{N^2}_{\ev}}\avg{N^2 \avg{(\tilde{x}_i{-}\beta_i \tilde{t})
        (\tilde{x}_j{-}\beta_j \tilde{t})}_S}_{\ev}, \label{R2ij_from_Cavg}
\end{equation}
where we suppressed the $\vec{K}$-dependence on the righthand side.  We refer to this way of computing the HBT radii as the ``average correlation function" method.  As an additional check on this result, we note that if we neglect event-by-event fluctuations entirely by setting $\delta S(x,K) = 0$, then $S(x,K) = \bar{S}(x,K)$, the final term in \eqref{Cavg_corrfunc_def} vanishes, and \eqref{R2ij_from_Cavg} reduces to \eqref{R2ij_from_Sbar}, as expected.  The correct theoretical definition of the ensemble-averaged HBT radii may therefore be thought of as a simple (weighted) average of the HBT radii for each fluctuating event, each scaled by a factor which accounts explicitly for final-state multiplicity fluctuations, bin by bin in $\vec{K}$, from event to event.

\subsubsection{Direct ensemble average}
\label{sec2b4}

A more direct route skips the construction of the correlation function entirely, and simply averages the radii \eqref{svHBT_defn}, computed from $S(x,K)$ event-by-event, directly:
\begin{equation}
\avg{R^2_{ij}}_{\ev}\!(\vec{K}) \equiv \frac{1}{N_{\ev}} \sum^{N_{\ev}}_{i=1} \l( R^2_{ij} \r)^{(k)}\!(\vec{K}), \label{R2ij_from_DEA}
\end{equation} 
where $\l( R^2_{ij} \r)^{(k)}$ denotes the HBT radii of the $k$th fluctuating event.  (Of course, this could also be done if the radii $\l( R^2_{ij} \r)^{(k)}$ were extracted from a Gaussian fit to the correlation function for event $k$ using Eq.~\eqref{corrfunc_functional_defn} with the corresponding emission function $S^{(k)}(x,K)$.  Here, however, $\l( R^2_{ij} \r)^{(k)}$ will be computed via the shortcut \eqref{svHBT_defn}, with $S$ replaced by $S^{(k)}$.)  This ``direct ensemble average" is clearly equivalent to the arithmetic mean of the event-by-event radii, without the additional multiplicity weight in \eqref{true_R2ij_from_corrfunc}.  Hereafter, we will drop the subscript '$\ev$' in the interest of notational simplicity, whenever doing so is unambiguous.

\section{Azimuthally averaged HBT interferometry for fluctuating sources}
\label{sec3}
\subsection{Azimuthally averaged HBT interferometry for a single event}
\label{sec3a}

Azimuthally sensitive HBT interferometry relies fundamentally on the construction of the two-particle correlation function given in Eq.~\eqref{corrfuncSEdefn}.  Again we first study the situation for a single event.  Taking \eqref{corrfuncSEdefn} as a starting point, there are at least two distinct ways of obtaining azimuthally averaged HBT radii.  The first is to construct the full azimuthally dependent correlation function, obtaining the azimuthally sensitive HBT radii by fitting \eqref{corrfuncSEdefn} to the form \eqref{corrfunc_functional_defn}, and then average explicitly over the residual dependence on $\Phi_K$:
\begin{eqnarray}
R^2_{ij,0}(K_T) &\equiv & \frac{1}{2\pi} \int^{\pi}_{-\pi} d\Phi_K R^2_{ij} \l( K_T, \Phi_K \r) \nonumber\\
	& \equiv & \azavg{R^2_{ij} \l( K_T, \Phi_K \r)}. \label{azavg_v1}
\end{eqnarray}

The second way to obtain azimuthally averaged HBT radii is to perform the average at the level of the correlation function \eqref{corrfuncSEdefn} before fitting HBT radii to it, instead of averaging the HBT radii after fitting the correlation function event by event.  Since the correlator is constructed as the ratio of two experimental quantities which are measured on a bin-by-bin basis, binning only on $K_T$ without binning on $\Phi_K$ avoids reducing the number of available particle pairs per bin by a factor of the number of bins in $\Phi_K$.  Theoretically, a correlator constructed in this way should be written as the ratio of the $\Phi_K$-averaged two-particle cross section divided by an uncorrelated background which is constructed by taking a product of the corresponding $\Phi_K$-averaged single-particle spectra:
\begin{equation}
\azavg{C}(\vec{p}_1, \vec{p}_2) = \frac{ \azavg{E_{p_1} E_{p_2} \frac{d^6 N}{d^3 p_1 d^3 p_2}}}{ \azavg{\l( E_{p_1} \frac{d^3 N}{d^3 p_1}\r)}\azavg{ \l(  E_{p_2} \frac{d^3 N}{d^3 p_2} \r) }}. \label{azavgcorrfuncSEdefn2}
\end{equation}
In terms of the emission function $S(x,K)$, this correlator may be written as
\begin{eqnarray}
\avg{C}_{\Phi_K}(\vec{q}, \vec{K})
	&\approx & \frac{\azavg{ \l| \int d^4x S(x,K) \r|^2}}{\l| \azavg{ \int d^4x S(x,K)} \r|^2} \label{corrfuncSEazavgdefn}\\
	&& \times	\l( 1 + \frac{\azavg{ \l| \int d^4x \e^{i q \cdot x} S(x,K) \r|^2}}{\azavg{ \l| \int d^4x S(x,K) \r|^2}} \r) \nonumber
\end{eqnarray}  
As before, we consider fitting this expression to the form \eqref{corrfunc_functional_defn} and extracting the $R^2_{ij}$ as fit parameters which depend only on $K_T$.  However, since the factor inside the parentheses in \eqref{corrfuncSEazavgdefn} tends to 2 in the limit that $q \rightarrow 0$, we must include an overall factor when fitting the correlator:
\begin{equation}
\azavg{C}\!(\vec{q},\vec{K}) \equiv C_0 \l( \! 1{+}\exp\!\l(-\!\sum_{i,j=o,s,l} q_i q_j R^{2}_{ij}(K_T)\r)\!\r)\!,\label{corrfuncSEazavgFITwithFACTORdefn}
\end{equation} 
where
\begin{equation}
C_0 \equiv \frac{\azavg{N^2(K_T, \Phi_K)}}{\azavg{N(K_T, \Phi_K)}^2},  \label{SEcorrnormFACTOR}
\end{equation}
with $N(K_T, \Phi_K) \equiv \int d^4x S(x,K)$ as before.  The azimuthally averaged $R^2_{ij}(K_T)$ are proportional to the curvature of the correlator at the origin, and one can show that this leads (with the functional dependence of $S(x,K)$ suppressed) to
\begin{eqnarray}
R^2_{ij} (K_T)
&=& \frac{\azavg{\int d^4x\, (x_i{-}\beta_i t)(x_j{-}\beta_j t)S}}{\azavg{\int d^4x\, S}} 
\nonumber\\
&-& \frac{\azavg{\int d^4x\, (x_i{-}\beta_i t)S}\azavg{\int d^4x\, (x_j{-}\beta_j t)S}}{\azavg{\int d^4 x S}^2} 
\nonumber\\
&=& \frac{\azavg{N^2(K_T, \Phi_K) R^2_{ij}(K_T, \Phi_K)}}{\azavg{N^2(K_T, \Phi_K)}} \label{azavg_v2}
\end{eqnarray}  
For Gaussian sources with azimuthally symmetric particle emission, the definition \eqref{azavg_v2} is equivalent to \eqref{azavg_v1}, i.e., $R^2_{ij}(K_T) = R^2_{ij,0}(K_T)$, since the factor $N^2(K_T, \Phi_K)$ is $\Phi_K$-independent and thus drops out from the ratio in \eqref{azavg_v2}.  For events with a significant azimuthal asymmetry in pair production, however, these two methods of azimuthal averaging can yield substantially different results.

Note that the prefactor outside the parentheses in \eqref{corrfuncSEazavgdefn} is independent of $q$ and thus also modulates the correlation function at $\l| \vec{q} \r| \rightarrow \infty$.  It does not affect the extraction of the HBT radii if, as is often done in experiment, the correlation function is normalized by hand to 1 at $\l| \vec{q} \r| \rightarrow \infty$.

\subsection{Azimuthally averaged HBT interferometry for ensembles of fluctuating events}
\label{sec3b}
\subsubsection{Azimuthally averaged HBT radii $R^2_{ij,0}(K_T)$}
\label{sec3b1}

In the previous subsection, we introduced two different ways of defining the azimuthally averaged HBT radii for a single event.  Analogously, for an ensemble of events, there are two different ways to average over the $\Phi_K$-dependence of each of the ensemble averaging methods defined in Eqs.~\eqref{R2ij_from_SSH}, \eqref{R2ij_from_Sbar}, \eqref{R2ij_from_Cavg} and \eqref{R2ij_from_DEA}.  We can either first perform an azimuthally sensitive HBT analysis, extract the $\Phi_K$-independent HBT radii, and average these over $\Phi_K$, or perform the $\Phi_K$-average already at the level of constructing the correlator (see Eq.~\eqref{azavgcorrfuncSEdefn2}) and then extract $\Phi_K$-independent radii from the azimuthally averaged correlator.  The first procedure requires higher event statistics, and is therefore experimentally more difficult.  Still, it is of conceptual interest and will thus be studied in this subsection.  The second method is experimentally preferred and will be discussed in the following subsection.  The ``direct ensemble average" of the azimuthally averaged HBT radii \eqref{azavg_v1} is given by
\begin{equation}
\avg{R^2_{ij,0}} \equiv \evavg{R^2_{ij,0}(K_T)} \equiv \azavg{\evavg{R^2_{ij} \l( K_T, \Phi_K \r)}}. \label{azavg_R2ij_from_DEA}
\end{equation} 
Since the azimuthal average commutes with the arithmetic average over events, $\avg{R^2_{ij}}$ is also the $\Phi_K$-average of the ensemble-averaged azimuthally symmetric radius \eqref{R2ij_from_DEA}.  Similarly constructed azimuthal averages of Eqs.~\eqref{R2ij_from_SSH}, \eqref{R2ij_from_Sbar} and \eqref{R2ij_from_Cavg} define their $\Phi_K$-independent parts, with
\begin{equation}
R^2_{\avg{ij},0}(\vec{K}) \equiv \azavg{R^2_{\avg{ij}}(\vec{K})}, \label{azavg_R2ij_from_Cavg}
\end{equation}
\begin{equation}
R^2_{\bar{ij},0}(\vec{K}) \equiv \azavg{R^2_{\bar{ij}}(\vec{K})}, \label{azavg_R2ij_from_SSH}
\end{equation} and
\begin{equation}
\bar{R}^2_{ij,0}(\vec{K}) \equiv \azavg{\bar{R}^2_{ij}(\vec{K})}. \label{azavg_R2ij_from_Sbar}
\end{equation}

\subsubsection{HBT radii $R^2_{ij}(K_T)$ from azimuthally averaged correlators}
\label{sec3b2}

In the same way that our treatment of azimuthally sensitive correlation functions in Sec.~\ref{sec2} admitted several different strategies for generalizing to ensemble-averaged correlators, the correlator \eqref{azavgcorrfuncSEdefn2} possesses several analogous ensemble-averaged generalizations.

We define the ``direct ensemble average" of the azimuthally averaged HBT radii by
\begin{eqnarray}
\avg{R^2_{ij}} &\equiv & \evavg{R^2_{ij}(K_T)} \nonumber\\
 	&\equiv & \evavg{\frac{\azavg{N^2(K_T, \Phi_K) R^2_{ij}(K_T, \Phi_K)}}{\azavg{N^2(K_T, \Phi_K)}}} \label{azavgDEA}
\end{eqnarray}
in accordance with \eqref{azavg_v2}.  In the cases of the ``single-shot hydrodynamics" and ``averaged emission function" methods, both approaches can be characterized by a single emission function (either $S_{{\mathrm{ssh}}}$ or $\bar{S}$), and therefore imply the following ensemble averaged generalizations of \eqref{azavg_v2}:
\begin{eqnarray}
R^2_{\bar{ij}}(K_T) \equiv \frac{\azavg{N_{{\mathrm{ssh}}}^2(K_T, \Phi_K) R^2_{\bar{ij}}(K_T,\Phi_K)} }{\azavg{N_{{\mathrm{ssh}}}^2(K_T, \Phi_K)}}  \label{azavgSSH}
\end{eqnarray} 
and 
\begin{eqnarray}
\bar{R}^2_{ij}(K_T) \equiv \frac{\azavg{\evavg{N(K_T, \Phi_K)}^2 \bar{R}^2_{ij}(K_T,\Phi_K)} }{\azavg{\evavg{N(K_T, \Phi_K)}^2}}, \label{azavgCbar}
\end{eqnarray} 
where $N_{{\mathrm{ssh}}}(K_T, \Phi_K)$ and $\evavg{N}(K_T, \Phi_K)$ are defined in an obvious way.

An equally straightforward, but slightly more tedious derivation shows that the ``average correlation function" method discussed in Sec. \ref{sec2b3}, which simulates the procedure applied in experimental analyses \cite{Lisa:2000xj,Adams:2003ra,Adare:2014vax}, can be adapted to the azimuthally averaged case by defining
\begin{eqnarray}
R^{2}_{\avg{ij}}(K_T) &=& \frac{\azavg{\evavg{N^2(K_T, \Phi_K) R^2_{ij}(K_T, \Phi_K)}}}{\azavg{\evavg{N^2(K_T, \Phi_K)}}}, \nonumber\\
	&=& \frac{\evavg{R^2_{ij}(K_T)\azavg{N^2(K_T, \Phi_K)}}}{\evavg{\azavg{N^2(K_T, \Phi_K)}}}.  \label{azavgCavg}
\end{eqnarray} 
Note that event-by-event fluctuations of the spectrum $N(K_T, \Phi_K)$ only enter in this last method which reproduces the experimental procedure.

\section{Comparison of theoretical ensemble averaging procedures}
\label{sec4}

\begin{figure}
\includegraphics[width=\linewidth]{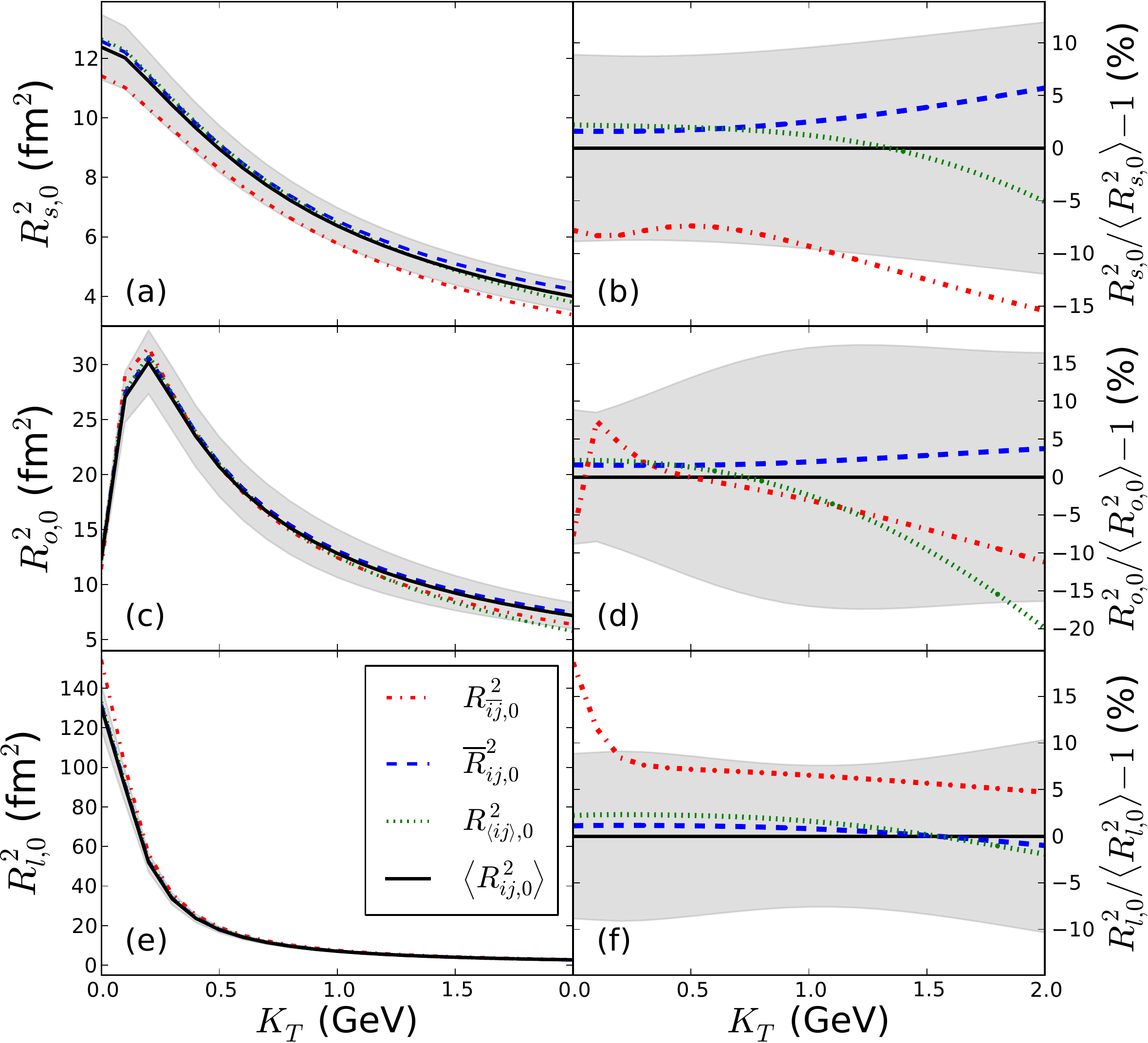}
\caption{The radius parameters extracted according to the four ensemble averaging methods described in Eqs.~\eqref{azavg_R2ij_from_DEA} - \eqref{azavg_R2ij_from_Sbar} (left column), as well as their percentage deviation from the direct ensemble average $\evavg{R^2_{ij,0}(K_T)}$ (right column), as functions of $K_T$.  The shaded bands represent the standard uncertainty of the direct ensemble average resulting from event-by-event fluctuations.
\label{Fig1:R2ijcfs0}}
\end{figure}

In general, each of the ensemble averaging methods discussed above yields different results.  While experimentally only the ``averaged correlation function" method is available, leading to the two possible ways \eqref{azavg_R2ij_from_Cavg} and \eqref{azavgCavg} to measure the HBT radii, the ``single-shot hydrodynamics" (Eqs.~\eqref{azavg_R2ij_from_SSH} and \eqref{azavgSSH}) and ``ensemble-averaged emission function" methods (Eqs.~\eqref{azavg_R2ij_from_Sbar} and \eqref{azavgCbar}) can be used in theoretical studies and offer significant numerical advantages.  It is therefore of interest to evaluate the significance of the differences between the different ensemble-averaging prescriptions.  We begin by comparing in Fig. \ref{Fig1:R2ijcfs0} the HBT radii $\avg{R^2_{ij,0}}$, $R^2_{\avg{ij},0}$, $\bar{R}^2_{ij,0}$ and $R^2_{\bar{ij},0}$ (defined in Sec. \ref{sec3b1}, Eqs.~\eqref{azavg_R2ij_from_DEA} - \eqref{azavg_R2ij_from_Sbar}) for each of the four prescriptions, applied to a typical hydrodynamic analysis using the iEBE-VISHNU package \cite{Shen:2014vra}.  Here, we consider 200 $A$ GeV Au+Au collisions at 0-10\% centrality, using the MC-Glauber model with p+p multiplicity fluctuations to compute the fluctuating initial entropy density profiles in the transverse plane, evolving them with boost-invariant hydrodynamics (with $\eta/s = 0$, i.e., assuming ideal fluid behavior) to simulate the evolution of the fireball.%
\footnote{%
	Taking $\eta/s{\,\neq\,}0$ tends to suppress the effects of event-by-event fluctuations,
	leading to much smaller discrepancies between the various methods of ensemble 
        averaging \cite{Plumberg:2015eia}.
	The corresponding discrepancies amongst the radii derived from ideal hydrodynamics
	(shown in the figures) therefore represent an upper limit on the extent to which the
	different methods of ensemble averaging may disagree with one another for arbitrary
	values of $\eta/s$.
}
We use $\nev = 5000$ which is large enough such that the observed variance of the HBT radii is dominated by event-by-event fluctuations, and fluctuations from finite sampling statistics can be neglected.  We terminate the hydrodynamical evolution along a freeze-out surface of constant temperature $T_{\mathrm{dec}} = 120$ MeV and use the Cooper-Frye algorithm to compute the charged particle yields.  Additional details of our analysis, as well as a more systematic discussion of the effects of shear viscosity on HBT analyses, are described in \cite{Plumberg:2015eia}.  Fig. \ref{Fig1:R2ijcfs0} shows that single-shot hydrodynamics, $R^2_{\bar{ij},0}$, (red dash-dotted line) leads to the least reliable theoretical estimates for the directly ensemble-averaged HBT radii $\avg{R^2_{ij,0}}$.  This reflects the strongly non-linear hydrodynamic response to event-by-event fluctuations in the initial density profile which single-shot hydrodynamics does not capture.  Both the ``average emission function" $\bar{R}^2_{ij,0}$ (blue dashed line) and ``average correlation function" $R^2_{\avg{ij},0}$ (green dotted line) methods yield results that are in much better agreement with the direct ensemble average $\evavg{R^2_{ij,0}}$ (black solid line), although they also tend to deviate from it at $K_T > 1$ GeV.

\begin{figure}
\includegraphics[width=\linewidth]{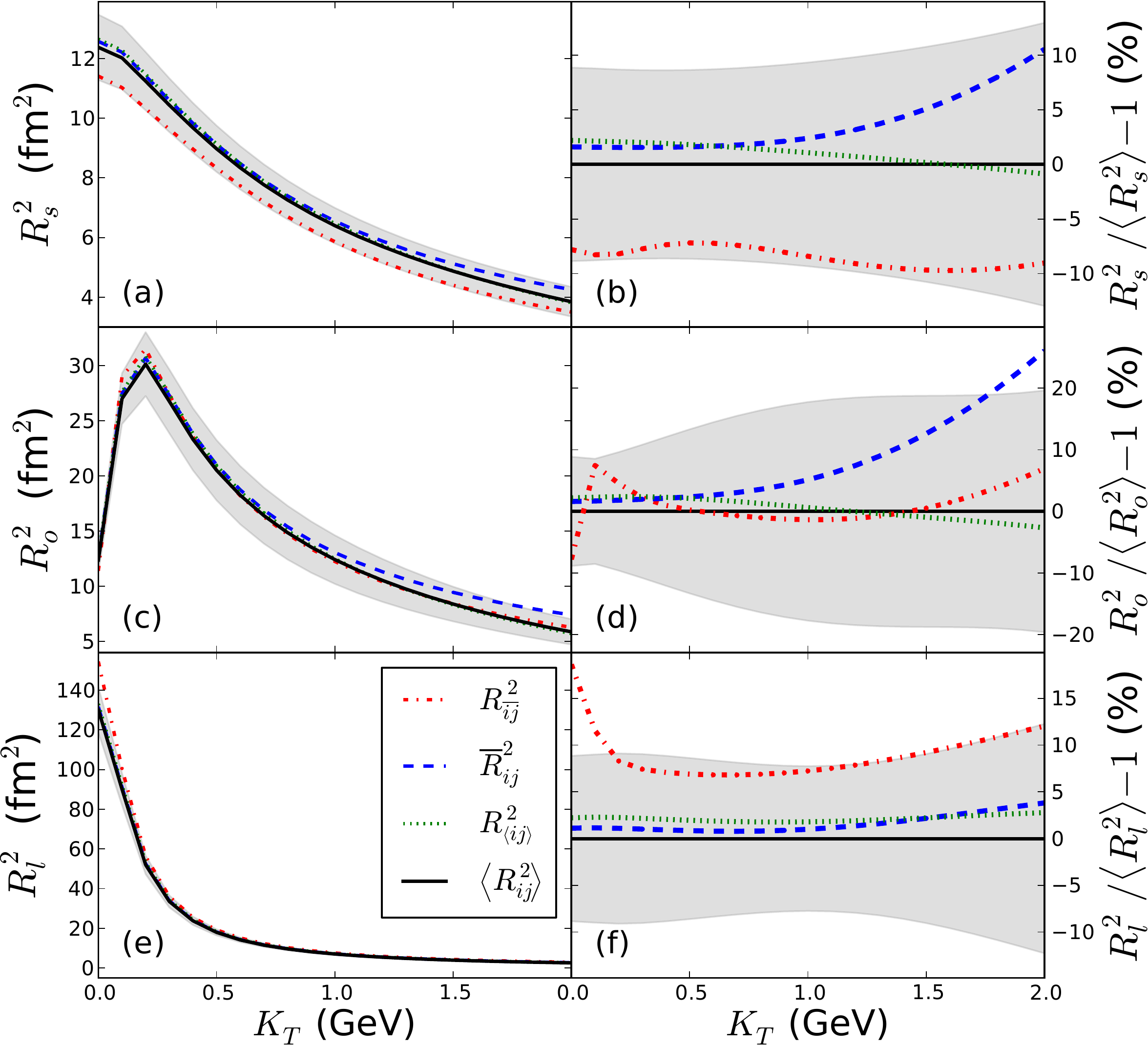}
\caption{ Similar to Fig. \ref{Fig1:R2ijcfs0}, but using the definitions \eqref{azavgDEA}-\eqref{azavgCavg} for the azimuthally and ensemble-averaged HBT radii.  
\label{Fig2:R2ijazavg}}
\end{figure}

In Fig.~\ref{Fig2:R2ijazavg} we show what happens to these four types of azimuthally and ensemble-averaged radii if the azimuthal average is not performed at the end of the HBT analysis on the level of the HBT radii, but instead on the level of constructing the correlation function, before extracting the HBT radii, as discussed in Sec. \ref{sec3b2}, Eqs.~\eqref{azavgDEA} - \eqref{azavgCavg}.  We focus our attention on the black solid and green dotted curves, representing the algebraic mean and experimentally determined average radii, $\avg{R^2_{ij}}$ and $R^2_{\avg{ij}}$, respectively.  We see that using the appropriate prescription for $\Phi_K$-averaging that applies to each method significantly improves the agreement between the experimentally accessible HBT radii $R^2_{\avg{ij}}$ and the theoretically interesting algebraic means $\avg{R^2_{ij}}$, especially at large $K_T$, when compared to the radii defined via Eqs.~\eqref{azavgDEA}-\eqref{azavgCavg} that were shown in Fig.~\ref{Fig1:R2ijcfs0}.

After this discussion of the first moment (i.e., the mean) of the event-by-event distribution of HBT radii, we now proceed to a discussion of the moments of this distribution.

\section{Estimating the direct ensemble average of the HBT radii}
\label{sec5}

In the preceding sections, we saw that the ensemble of events is characterized by an event-by-event distribution of HBT radii $R^2_{ij}$, and that the experimentally extracted ensemble-averaged HBT radii track, but do not exactly reproduce the mean value of this distribution.  In this section, we describe a way of experimentally estimating this mean (i.e., the direct ensemble average $\avg{R^2_{ij}}$), using only the experimentally accessible weighted sample averages from this distribution. We present this method as for a general observable $\mcO$ that can be defined on an event-by-event basis, although in this paper we will eventually restrict its application to the HBT radii.  We will refer to this method as ``mean estimation."

For a general physical observable $\mcO$ that is defined on an event-by-event basis and fluctuates from event to event, we distinguish between two types of distributions: (i) the underlying physical probability distribution $\mcP(\mcO)$ which, for continuous $\mcO$, is in general a continuous function, and (ii) the discrete distribution $\mcP_n(\mcO)$ that describes the distribution of values $\mcO_1,\ldots,\mcO_n$ obtained in $n$ measurements of the observable $\mcO$.  As $n \rightarrow \infty$, the discrete distribution $\mcP_n(\mcO)$, properly normalized, approaches the physical distribution $\mcP(\mcO)$.  We will refer to $\mcP_n(\mcO)$ as the ``measured distribution of $\mcO$" or the ``ensemble distribution of $\mcO$" in our ensemble of $n$ measured events, while $\mcP(\mcO)$ will be called the ``true" or ``physical" distribution of $\mcO$.

For an arbitrary distribution $\mathcal{Q}(\mcO)$, we also define for later use the associated distribution $\bar{\mathcal{Q}}_n(\mcO)$ as the distribution of sample means of observable $\mcO$ of size $n$ sampled from a physical distribution $\mathcal{Q}(\mcO)$.  Thus, the distribution of sample means of size $n$ from the true physical distribution $\mathcal{P}(\mcO)$ above is denoted by $\bar\mcP_n(\mcO)$.  For a measured distribution $\mcP_N(\mcO)$ of size $N$ we denote the analogous distribution of sample means of size $n<N$ by $\bar\mcP_{N,n}(\mcO)$.  For sufficiently large ensembles (i.e., as $N \rightarrow \infty$), $\bar\mcP_{N,n}(\mcO)$ converges to $\bar\mcP_{n}(\mcO)$, in the same way that $\mcP_N(\mcO)$ converges to $\mcP(\mcO)$.  Our goal will be to estimate moments of the ensemble distribution $\mcP_N(\mcO)$ (and thereby, the physical distribution $\mathcal{P}(\mcO)$) by studying the statistical moments of $\bar\mcP_{N,n}(\mcO)$ in repeated sets of samplings of size $n$, in the limit of sufficiently large $N$.

We now show how to estimate the direct ensemble average $\avg{\mcO}$ of an observable $\mcO$.  We reiterate that this quantity, in the context of HBT interferometry, cannot be directly measured experimentally, since single heavy-ion collisions yield total multiplicities which are too small for extracting meaningful estimates of the radii event by event.  The fundamental observables available from experimental HBT analyses are therefore only multiplicity-weighted averages of the HBT radii as defined in Eq.~\eqref{true_R2ij_from_corrfunc}.  We thus consider weighted averages of the observable $\mcO$ of the form
\begin{equation}
\avg{w\mcO}_N \equiv \sum^N_{k=1} w^{(N)}_k \mcO_k,  \label{general_weighted_average}
\end{equation} 
where $k$ is an index running over all $N$ events, and the weights are subject to the following normalization condition:
\begin{equation}
\sum^N_{k=1} w^{(N)}_k = 1.   \label{weight_normalization_conditions}
\end{equation}  
For instance, Eq.~\eqref{true_R2ij_from_corrfunc} may be obtained by taking $N=\nev$, $\mcO = R^2_{ij}(\vec{K})$, and $w^{(\nev)}_k = N^2_k(\vec{K}) / \sum^{\nev}_{k=1} N_k^2(\vec{K})$.  We now show how to construct estimates for the moments of the event-by-event distribution $\mcP(\mcO)$ in terms of expressions of the form \eqref{general_weighted_average}.

In order to estimate the direct ensemble average
\begin{equation}
\avg{\mcO} = \lim_{N \rightarrow \infty} \frac{1}{N} \sum^{N}_{k=1} \mcO _k \label{defn_of_true_mean}
\end{equation} 
of $\mcO$ in terms of the weighted average $\avg{w \mcO}_N$ in Eq.~\eqref{general_weighted_average} we must find a way to correct for the weights $w^{(N)}_k$ that are an unavoidable part of the experimental measurement.  For the purpose of this paper we will assume that the weights $w^{(N)}_k$ are measurable event by event (such as the multiplicity weights above).  We will also assume that for every value $w$ of the weight our measured sample contains many events with weights $w^{(N)}_k$ close to $w$.

As a first step, let us sort the events by increasing weight $w^{(N)}_k$.  Next, we create $n_b$ bins and fill each bin with the same number $n = N/n_b$ in order of increasing weight $w^{(N)}_k$.  We label these bins by $(\ell)$, $\ell = 1, \dots, n_b$.  For each bin $(\ell)$ we construct the weighted average
\begin{equation}
\avg{w \mcO}^{(\ell)}_n \equiv \sum_{k \in (\ell)} w^{(n)}_{k} \mcO_k
\end{equation} 
with the modified weights
\begin{equation}
w^{(n)}_{k} \equiv \frac{w^{(N)}_{k}}{\sum_{k \in (\ell)} w^{(N)}_{k}}. \label{modified_weights}
\end{equation}  
$\avg{w \mcO}^{(\ell)}_{n}$ is a weighted average of the type that can be measured experimentally (such as the HBT radii \eqref{true_R2ij_from_corrfunc}).  If the number $n$ of events in each bin is large enough, this weighted average will be known with good statistical precision, i.e. it will closely approximate the corresponding weighted average of the underlying physical distribution $\mcP({\mcO})$.  If $n$, while being large, is much smaller than the total number of events $N$, the monotonically increasing modified weights $w^{(n)}_{k}$ will not show much variation%
\footnote{%
	If one of the bins happens to contain events from a stretch in the ordered list where
	$w^{(n)}_{k}$ rises abruptly, and thus the condition of small variation of the
	weights inside that bin is violated, this bin may be thrown away.  In essence, the
	method described in this section corresponds to binning the total ensemble of 
        events into bins with approximately the same weight (multiplicity) such that within 
        the bin weight
	(multiplicity) fluctuations can be neglected, and then averaging the bin averages over
	all bins.  If total statistics in the ensemble is sufficient, this last average over
	weight bins need not be performed, and this would allow us in our case to
	study whether events with different multiplicities have different mean HBT radii.
}
within each bin $(\ell)$ and will all have approximately the same magnitude $1/n$:
\begin{equation}
w^{(n)}_{k} \approx \frac{1}{n} = \frac{n_b}{N}
\end{equation}  
We can now arithmetically average the weighted bin averages to find
\begin{eqnarray}
\frac{1}{n_b} \sum^{n_b}_{\ell=1} \avg{w \mcO}^{(\ell)}_{n}
	&=& \frac{1}{n_b} \sum^{n_b}_{\ell=1} \sum_{k \in (\ell)} w^{(n)}_{k} \mcO_k \nonumber\\
	&\approx & \frac{1}{n_b} \sum^{n_b}_{\ell=1} \sum_{k \in (\ell)} \frac{n_b}{N} \mcO_k 
\nonumber\\
	&=& \frac{1}{N} \sum^{N}_{k=1} \mcO_k 
\nonumber\\
	&\equiv & \avg{\mcO}_N  
\label{bin_average_approximate_mean}
\end{eqnarray}  
The approximation in the second line above becomes exact as $N, n_b \rightarrow \infty$.  The ensemble average $\avg{\mcO}_N$ is the mean of the ensemble distribution $\mcP_N(\mcO)$ mentioned above.  For large $N$ it approaches the true mean $\avg{\mcO}$ of the observable:
\begin{equation}
\lim_{N \rightarrow \infty} \avg{\mcO}_N = \lim_{N,n_b \rightarrow \infty} 
\frac{1}{n_b} \sum^{n_b}_{\ell=1} \avg{w \mcO}^{(\ell)}_{n} = \avg{\mcO}.
\end{equation}  
In Sec.~\ref{sec8} we will show numerical results for a toy example where taking $n_b \geq 10$ in \eqref{bin_average_approximate_mean} is sufficient to obtain an estimate of $\avg{\mcO}_N$ with statistical error of less than 1\%.

\section{Extracting the scale of fluctuations in the HBT radii}
\label{sec6}
\subsection{Estimating the variance}
\label{sec6a}

We now proceed to show how to estimate the variance of the event-by-event distribution of an observable $\mcO$.%
\footnote{%
	Note that our use of the term ``variance estimation" in this paper differs from its more
	common usage in the field of statistics to refer to the general set of techniques for
        gauging the precision of estimators derived from sample data.  Although based on
	similar principles, the method of variance estimation presented here bears only a
 	superficial resemblance to this set of techniques.  For further discussion see, for
 	instance, \cite{Wolter:2007itve}.
}

The variance of the ensemble distribution of $\mcO$ is defined by%
\footnote{%
	The factor of $1/(N{-}1)$ is used in this expression in place of $1/N$ in order to
	render it an \textit{unbiased} estimator \cite{Tamhane:2000sda} of the variance of
	the physical distribution of $\mcO$.
}
\begin{equation}
\sigma^2_{\mcO,N}
	\equiv \var{\mcP_N(\mcO)}
	\equiv \frac{1}{N-1} \sum_{i=k}^{N} \l( \mcO_k^2 - \avg{\mcO}_N^2 \r), \label{defn_of_true_variance}
\end{equation}
where $\avg{\mcO}_N$ is the ensemble mean of $\mcO$ defined in \eqref{bin_average_approximate_mean}.  The variance of the physical distribution of $\mcO$ is then related to $\sigma^2_{\mcO,N}$ by
\begin{equation}
\sigma^2_{\mcO} = \lim_{N \rightarrow \infty} \sigma^2_{\mcO,N}.	\label{ensemble_and_physical_variances_relation}
\end{equation}

To estimate $\sigma^2_{\mcO,N}$, we assume a very large ensemble of $N$ events and consider the process of randomly splitting this ensemble into $n_b$ bins of $n \equiv N / n_b$ events each, computing the \textit{sub}-ensemble average $\avg{\mcO}_n$ of $\mcO$ for each bin.  For many different repetitions of this process, the entire collection of sub-ensemble averages obtained yields the distribution of sample averages of $\mcO$ over sub-ensembles of size $n$ from an ensemble containing $N$ events, introduced in Sec.~\ref{sec5} and denoted by $\bar\mcP_{N,n}(\mcO)$.  The central limit theorem guarantees that the variance of $\bar\mcP_{N,n}(\mcO)$ is proportional to the variance of the ensemble distribution of $\mcO$, $\mcP_N(\mcO)$:
\begin{equation}
\var{\bar\mcP_{N,n}(\mcO)} \propto \var{\mcP_N(\mcO)}.
\end{equation}  
In the limit $N \rightarrow \infty$, the proportionality factor is simply $1/n$; for finite $N$, an additional finite population correction factor must be included (see Sec.~\ref{sec7}).

To formulate an explicit estimate for the variance of the ensemble distribution of $\mcO$, let $M$ be the total number of times the ensemble of size $N$ is split into $n_b$ bins of size $n$, and let $\avg{\mcO}_{k,\ell}$ represent the sub-ensemble average of $\mcO$ in the $\ell$th bin of the $k$th binning iteration (we suppress the dependence of $\avg{\mcO}_{k,\ell}$ on the bin size $n$ to reduce clutter).  Then we can show (see appendix~\ref{App:AppendixA}) that
\begin{equation}
\sigma^2_{\mcO, N, {\mathrm{est}}} \equiv \frac{N}{M n_b(n_b-1)} \sum_{k=1}^M \sum_{\ell=1}^{n_b} \l( \avg{\mcO}_{k,\ell}^2 - \avg{\mcO}_N^2 \r) \label{defn_of_method}
\end{equation}  
is a variance estimator that converges to $\sigma^2_{\mcO,N}$ as $M$ approaches the maximal number of different binnings $M_{\mathrm{max}}(N,n_b)$ defined in \eqref{Mmaxdefn}. According to Eq.~\eqref{ensemble_and_physical_variances_relation}), $\sigma^2_{\mcO, N}$ itself converges to the variance $\sigma^2_{\mcO}$ of the underlying physical distribution $\mcP(\mcO)$ as $N \rightarrow \infty$. Again, we suppress the dependence of $\sigma^2_{\mcO, N, {\mathrm{est}}}$ on $n_b$ and $M$ for clarity. For the particular case of $\mcO \equiv R^2_{ij}$ (and defining  $\sigma^2_{ij} \equiv \sigma^2_{R^2_{ij}}$), we have
\begin{equation}
\sigma^2_{ij, N, {\mathrm{est}}} \equiv \frac{N}{M n_b(n_b-1)} \sum_{k=1}^M \sum_{\ell=1}^{n_b} \l( \avg{R^2_{ij}}_{k,\ell}^2 - \avg{R^2_{ij}}_N^2 \r) \label{defn_of_R2ijmethod}
\end{equation} 

Each sub-ensemble average $\avg{\mcO}_{k,\ell}$ may be estimated by subdividing the corresponding bin of size $n$ into $\tilde{n}_b=n/\tilde{n}$ sub-bins of size $\tilde{n}$ and employing the method of mean estimation presented in the previous section. Thus, we are able to estimate the variance (or, equivalently, the \textit{second} central moment) of the ensemble distribution of an observable $\mcO$ by computing the distribution of sub-ensemble averages (i.e., \textit{first} moments) from the same distribution.  Furthermore, the equivalence between $\sigma^2_{\mcO,N,{\mathrm{est}}}$ and $\sigma^2_{\mcO,N}$ holds exactly in the limit $M \rightarrow M_{\mathrm{max}}(N,n_b)$.  However, even for moderate $N$, $M_{\mathrm{max}} = N! / (n!)^{n_b}$ is huge.  For example, $N=10$ and $n_b = 5$ yields $M_{\mathrm{max}} \sim 10^5$; for $N=100$ and $n_b=2$, $M_{\mathrm{max}} \sim 10^{29}$.  So in practice the limit $M \rightarrow M_{\mathrm{max}}$ is out of reach.  Fortunately, we will see that $\sigma^2_{\mcO,N,{\mathrm{est}}} \approx \sigma^2_{\mcO,N}$ to a very good approximation, even for $M \ll M_{\mathrm{max}}$.

Rather than summing over all possible distinct ways of sorting $N$ events into $n_b$ bins, one can thus evaluate \eqref{defn_of_method} by summing only over a sufficiently large number $M$ of subdivisions such that $\sigma^2_{\mcO,N,{\mathrm{est}}}$ converges to a fixed value within a given tolerance. In this way, the process of evaluating the righthand side of $\eqref{defn_of_method}$ can be performed cumulatively. Furthermore, each iteration of the sorting and bin-averaging procedures can be performed independently of all the others, implying that our method can be easily parallelized for numerical computation.

\subsection{Constructing the covariance matrix}
\label{sec6c}

The method discussed in subsection~\ref{sec6a} is readily extended to incorporate event-by-event fluctuations of multiple observables, and the correlations between them.  These correlations may be quantified by the covariance matrix between the observables of interest, and each element of this matrix may be estimated using a straightforward generalization of \eqref{defn_of_method}:
\begin{eqnarray}
&&{\cov{\mcO_1}{\mcO_2}}_{N,{\mathrm{est}}} \equiv \frac{N}{n_b (n_b - 1) M} \label{defn_of_3Dmethod}\\
	 && \times \sum^M_{k=1} \sum^{n_b}_{l=1} \l( \avg{\mcO_1}_{k,l}{-}\avg{\mcO_1}_N \r) \l( \avg{\mcO_2}_{k,l}{-}\avg{\mcO_2}_N \r). \nonumber
\end{eqnarray}  
For $\mcO_1 \equiv R^2_{ij}$ and $\mcO_2 \equiv R^2_{i'j'}$, this expression becomes
\begin{eqnarray}
&&{\cov{R^2_{ij}}{R^2_{i'j'}}}_{N,{\mathrm{est}}} \equiv \frac{N}{n_b (n_b - 1) M} \label{defn_of_3DR2ijmethod}\\
	&& \times \sum^M_{k=1} \sum^{n_b}_{l=1} \l( \avg{R^2_{ij}}_{k,l}{-}\avg{R^2_{ij}}_N \r)\!\l( \avg{R^2_{i'j'}}_{k,l}{-}\avg{R^2_{i'j'}}_N \r) .  \nonumber
\end{eqnarray}  
For the most general case in three dimensions, this leads to a $6 \times 6$ covariance matrix for the full set of $R^2_{ij}$. When $i'=i$ and $j'=j$, this expression reduces to a simple generalization of \eqref{defn_of_method} to three dimensions.%
\footnote{%
	For the discussions and derivations presented so far, we have written our results
	in terms of fluctuating radii $R^2_{ij}$.  Often in the literature, however, the reported
	quantities are not simply $R^2_{ij}$ but $R_{ij}$, so that one has a choice whether 
	to report properties of a distribution of the squared radii or the radii themselves.  
	In this paper we will adopt the former approach: the results we present are the
 	moments of the event-by-event distribution of the fluctuating $R^2_{ij}$.
}

\section{Generalization to higher moments}
\label{sec7}

We can generalize the combination of the methods introduced in sections \ref{sec5} and \ref{sec6} to permit the extraction of higher moments of $\mcP(\mcO)$.  This method, which we refer to as ``higher moment estimation," assumes the ability to estimate or calculate the direct ensemble average, and therefore relies on the method of mean estimation already discussed.  We first recast our estimate for the variance \eqref{defn_of_R2ijmethod} in terms of the second central moment of $\mcP_N(\mcO)$:%
\footnote{%
	For a smooth distribution, such as the true HBT distribution underlying all HBT
 	measurements, the variance is identically equal to the second central moment.  
	For the ensemble of measured events, however, the sample size $N$ is large but
 	finite, and the variance differs from the second central moment by a factor of 
	$\l( \frac{N-1}{N} \r)$ \cite{Tamhane:2000sda}.
}
\begin{eqnarray}
M_{2,N} & \equiv & \frac{1}{N} \sum^N_{k=1} \l( \mcO_k - \avg{\mcO}_N \r)^2 = 
\l( \frac{N{-}1}{N} \r) \sigma^2_{\mcO,N} 
\nonumber\\
	& \approx & \l(\frac{n(N{-}1)}{N{-}n}\r) 
\l( \frac{1}{n_b M}\sum^M_{j=1} \sum^{n_b}_{k=1} \l( \avg{\mcO}_{j,k}{-}\avg{\mcO}_N \r)^2 \r) \nonumber\\
	& \equiv & M_{2,N,{\mathrm{est}}}. 
\label{secondcentralmomentdefn}
\end{eqnarray}  
Here, we have introduced the notation $M_k$, $k\geq 2$, for the $k$th central moment of the measured $\mcO$ distribution $\mcP_N(\mcO)$, defined by
\begin{equation}
M_{k,N} \equiv \frac{1}{N} \sum^N_{k=1} \l( \mcO_k - \avg{\mcO}_N \r)^k. \label{kth_central_moment_defn}
\end{equation}  
The r.h.s. of the \eqref{secondcentralmomentdefn} consists of two factors: the first is a correction factor which accounts for the finite number of events $N$ in the ensemble, while the second factor is the second central moment of the distribution of bin-averages of size $n \equiv N/n_b$, sampled from the ensemble.  $\avg{\mcO}_{j,k}$ is a random variable $\mco$ distributed according to the distribution $\bar{\mcP}_{N,n}$ defined in Sec.~\ref{sec5}. Defining additionally $M_{1,N} \equiv \avg{\mcO}_N$, we can thus write
\begin{eqnarray}
M_{2,N} = \l(\frac{n(N{-}1)}{N{-}n}\r) \avg{\l( \mco - M_{1,N} \r)^2}_{N,n},
\end{eqnarray} 
where $\avg{\ldots}_{N,n}$ denotes the expectation value with respect to the distribution $\bar{\mcP}_{N,n}$.

The extension to higher order moments can be found in textbooks (e.g., \cite{Kendall:1987ast}). For the third and fourth order moments (related to the skewness and the kurtosis) one finds
\begin{eqnarray}
\avg{\l( \mco - M_{1,N} \r)^3}_{N,n} &=& \frac{(N{-}n)(N{-}2n)}{n^2(N{-}1)(N{-}2)}\, M_{3,N}  \\
\avg{\l( \mco - M_{1,N} \r)^4}_{N,n} &=& \frac{N{-}n}{n^3(N{-}1)(N{-}2)(N{-}3)} \\
	& &\times  \l[ \l( N^2{-}6nN{+}N{+}6n^2 \r) M_{4,N} \r. \nonumber\\
	& & \l.+3N(n{-}1)(N{-}n{-}1)\, M_{2,N}^2\r], 
\nonumber
\end{eqnarray} 
which can be solved for $M_{3,N}$ and $M_{4,N}$. Using our earlier notation we can thus write our estimates for the moments $M_{3,N}$ and $M_{4,N}$ as
\begin{eqnarray}
M_{3,N,{\mathrm{est}}} &=& \l( \frac{n^2(N{-}1)(N{-}2)}{(N{-}n)(N{-}2n)} \r) \nonumber\\
	& \times & \l( \frac{1}{n_b M}\sum^M_{k=1} \sum^{n_b}_{\ell=1} \l( \avg{\mcO}_{k,\ell}{-}\avg{\mcO}_N \r)^3 \r) \label{M3_est}
\end{eqnarray}
\begin{widetext}

\begin{eqnarray}
M_{4,N,{\mathrm{est}}} &=& \frac{1}{ N^2{-}6nN{+}N{+}6n^2} \BIGGL \l( \frac{n^3(N{-}1)(N{-}2)(N{-}3)}{N{-}n} \r) 
\l( \frac{1}{n_b M}\sum^M_{k=1} \sum^{n_b}_{\ell=1} \l( \avg{\mcO}_{k,\ell}{-}\avg{\mcO}_N \r)^4 \r)
\nonumber\\
& & \qquad \qquad \qquad - 3N(n{-}1)(N{-}n{-}1) \l(\frac{n(N{-}1)}{N{-}n}\r)^2 \l( \frac{1}{n_b M}\sum^M_{k=1} \sum^{n_b}_{\ell=1} \l( \avg{\mcO}_{k,\ell}{-}\avg{\mcO}_N \r)^2 \r)^2 \BIGGR 
\label{M4_est}
\end{eqnarray}
\end{widetext}

In terms of the central moments of a given distribution, we can also define its \textit{skewness} and \textit{excess kurtosis}:
\begin{eqnarray}
\beta_{3,N} &\equiv & \frac{M_{3,N}}{M_{2,N}^{3/2}}, 
\label{beta3_defn}\\
\beta_{4,N} - 3 &\equiv & \frac{M_{4,N}}{M_{2,N}^2} -3.   
\label{beta4m3_defn}
\end{eqnarray}  
We will find these definitions useful when we demonstrate the validity of our method below.  Eqs.~\eqref{beta3_defn} and \eqref{beta4m3_defn} may also be used to obtain $\beta_{3,N,{\mathrm{est}}}$ and $\beta_{4,N,{\mathrm{est}}}{-}3$, with each occurrence of $M_{k,N}$ replaced by $M_{k,N,{\mathrm{est}}}$ as defined above.

\section{Results and discussion}
\label{sec8}
\subsection{Estimating the arithmetic mean}
\label{sec8a}

In this section we present several proof-of-principle demonstrations of the various methods discussed in this paper.  In the present subsection, we illustrate the method presented in Sec.~\ref{sec5} to estimate the arithmetic average from particular linear combinations of weighted sub-averages. In Sec.~\ref{sec8b}, we discuss our method for estimating higher moments of a distribution by fleshing out the sampling distribution of the arithmetic average for the same distribution. Finally, in Sec.~\ref{sec8c}, we combine these two methods and use them in a more realistic scenario to estimate the relative variance of event-by-event fluctuations in the HBT radii from a sample of hydrodynamically evolved fireballs with fluctuating initial conditions.

We now illustrate our procedure for estimating the arithmetic average from weighted sub-averages.  To do this, we consider a joint binormal distribution of two random variables, which we label $X$ (the observable of interest) and $w$ (the unnormalized weight attached to each measurement):
\begin{eqnarray}
P(\mu_X, \mu_w, \sigma_X, \sigma_w, \rho; X,w) \equiv \quad \quad \quad \quad \quad \quad \quad \quad \quad \quad \\
\frac{\exp\l[ -\frac{1}{1-\rho^2} \l( \frac{(w-\mu_w)^2}{2\sigma_w^2}{+}\frac{(X-\mu_X)^2}{2\sigma_X^2}{-}\frac{\rho(w-\mu_w)(X-\mu_X)}{\sigma_w \sigma_X} \r) \r]}{2\pi \sqrt{1-\rho^2} \sigma_X \sigma_w} \nonumber
\end{eqnarray}
We treat these two variables as governed by a joint distribution with a non-trivial covariance matrix.

For specificity, we take the following set of parameters:
\begin{eqnarray}
\mu_X &\equiv& 10, \mu_w \equiv 7 \\
\sigma_X &\equiv& 3, \sigma_w \equiv 1/5 \\
\rho &\equiv & 0.0, 0.2, 0.4, 0.6, 0.8, 0.999
\end{eqnarray}  
The last variable, $\rho$, controls the correlation between the stochastic variables $X$ and $w$.  

\begin{figure}
\centering
\includegraphics[width=\linewidth]{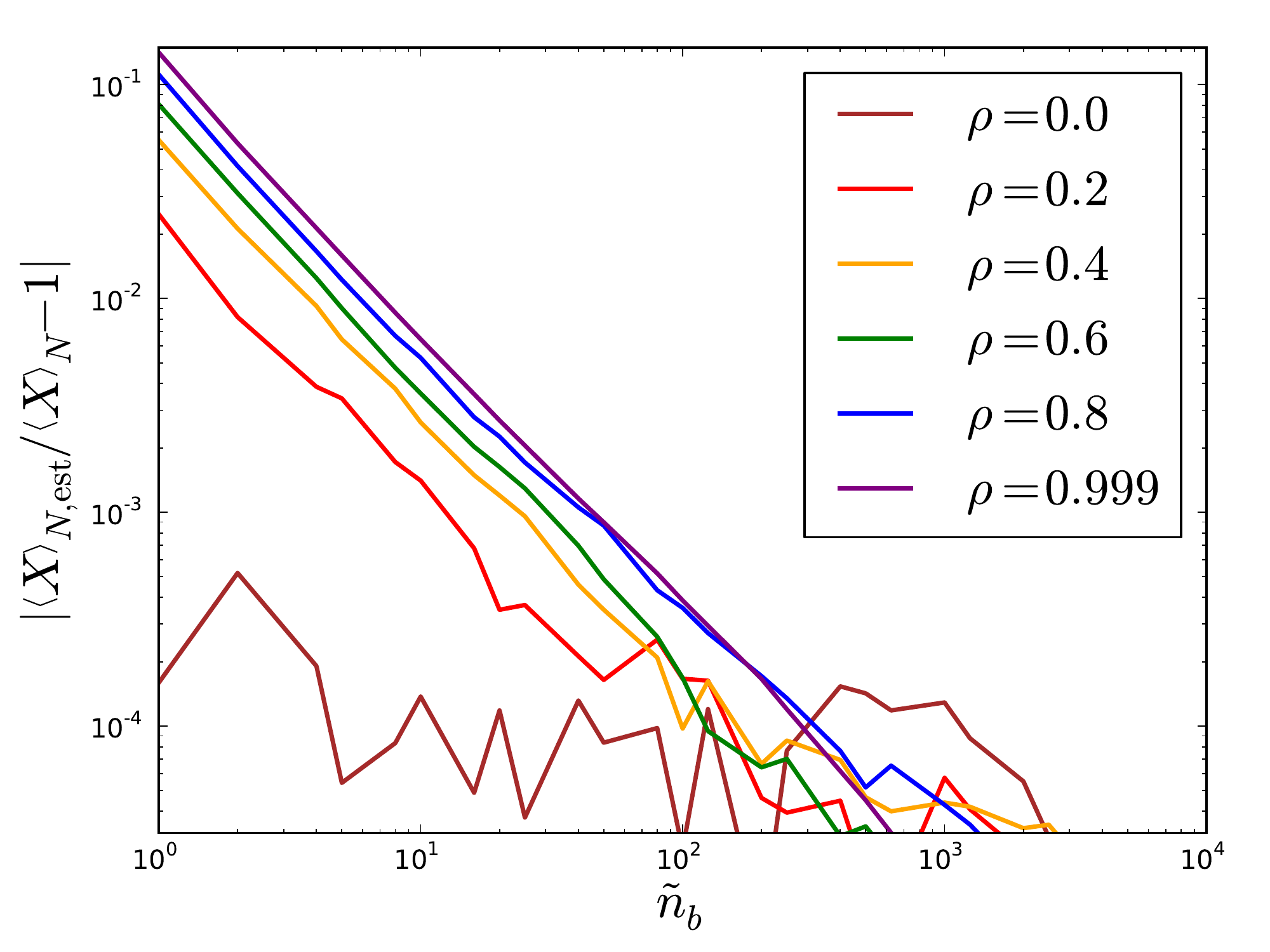} 
\caption{The estimated mean $\avg{X}_{N,\mathrm{est}}$, compared with the exact mean $\avg{X}_{N}$, as a function of $\tilde{n}_b$ and $\rho$.  We see that decreasing the strength of the correlation $\rho$ leads to a reduction in the overall relative difference between $\avg{X}_{N,\mathrm{est}}$ and $\avg{X}_{N}$, since total decorrelation requires that the weighted average factorizes and reduces to the arithmetic mean.  We note additionally that, for $\tilde{n}_b \gtrsim 10$, our method of estimating the arithmetic mean is accurate to better than 1\%, even with nearly total correlation between the $w_i$ and the $X_i$.
\label{Fig3:mean_vs_tnb_est}}
\end{figure}

We sample from this distribution $N=10,\!000$ observation-weight pairs $(X_i, w_i)$. Using the procedure discussed in Sec.~\ref{sec5}, we repeatedly distribute these $N$ events randomly into $\tilde{n_b}$ bins of size $\tilde{n}=N/\tilde{n}_b$ and estimate the arithmetic mean $\avg{X}_N$ from Eq.~\eqref{bin_average_approximate_mean}. The results are plotted in Fig.~\ref{Fig3:mean_vs_tnb_est} as a function of the number of bins $\tilde{n}_b$, for several different values of the correlation coefficient $\rho$.  We compute $\avg{X}_{N,\mathrm{est}}$ from the lefthand side of Eq.~\eqref{bin_average_approximate_mean} and compare it with the exact mean $\avg{X}_N$ of the $N$ sampled values. 

Figure~Fig.~\ref{Fig3:mean_vs_tnb_est} demonstrates clearly that  for $\tilde{n}_b \sim 10$ we can approximate the arithmetic average $\avg{X}_N$ to better than 1\%, as noted earlier.  All curves eventually drop to zero when $\tilde{n}_b = N = 10,\!000$, since this corresponds to placing each event in its own bin, so that the lefthand side of \eqref{bin_average_approximate_mean} reduces to the arithmetic average $\avg{X}_N$.
In between, i.e. for $10 < \tilde{n_b} < 10,\!000$, we note that the plotted ratio starts to vary irregularly as a function of $n_b$ once it reaches a level around $10^{-4}$, reflecting the fundamental granularity of the measured sample of $N$ ``events'' as explained below.  

In Sec.~\ref{sec3}, we noted that bins which contained abruptly changing weights after ordering should be discarded; here, however, we have included all bins for the sake of simplicity, regardless of how significantly the weights vary within a bin.  Fig.~\ref{Fig3:mean_vs_tnb_est} demonstrates that this choice does not impede our ability to reliably estimate the mean $\avg{X}_N$.

In the particular case of $\rho = 0$ and $N \rightarrow \infty$ (keeping $\tilde{n}_b$ finite), each weighted bin-average $\avg{w\mcO}^{\ell}$ on the lefthand side of \eqref{bin_average_approximate_mean} factorizes into a product of the average weight and average observable in each bin. Each weighted bin-average thus reduces to the corresponding arithmetic bin-average, causing the normalized difference between $\avg{X}_{N,\mathrm{est}}$ and $\avg{X}_N$, which is plotted in Fig.~\ref{Fig3:mean_vs_tnb_est}, to vanish. The fact that in 
Fig.~\ref{Fig3:mean_vs_tnb_est}  the curve for $\rho = 0$ does \textit{not} collapse identically to zero can be understood as the result of event-by-event fluctuations of the event weights about their average values in each bin. The presence of these fluctuations imposes a lower bound on the error in our method of mean estimation which in Fig.~\ref{Fig3:mean_vs_tnb_est} is of order $10^{-4}$.  We have found empirically that this lower bound on the error scales approximately as $O\bigl(1/\sqrt{N \tilde{n}_b}\bigr)$. The accuracy of the method therefore increases as the number of events in the ensemble $N$ and the number of bins $\tilde{n}_b$ employed in the method are increased.

\subsection{Estimating higher moments}
\label{sec8b}

%
\begin{figure}[b]
\centering
\includegraphics[width=\linewidth]{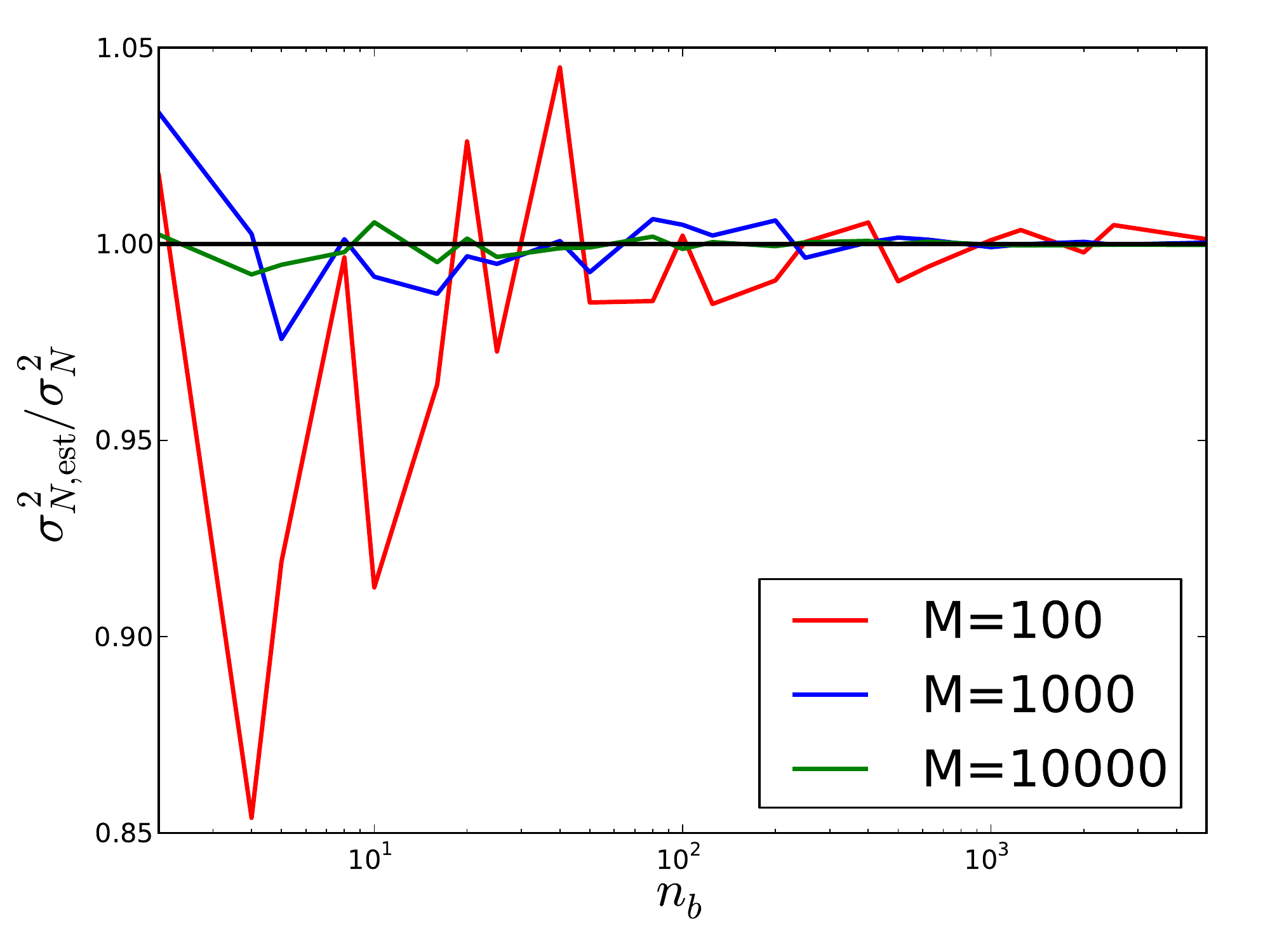} 
\caption{The estimated variance $\sigma^2_{N,\mathrm{est}}$, compared with the exact variance $\sigma^2_N$, for $N=10,\!000$, as a function of $n_b$ and M.
\label{Fig4:sig2_vs_nb_est}}
\end{figure}
%
To demonstrate the robustness of our method for extracting the arithmetic mean and using it to estimate higher moments of a distribution, we take a random sample of $N = 10,\!000$ observations from a skew normal distribution characterized by the following probability density function:
\begin{equation}
P(\mu,\sigma,\alpha; X) \equiv \frac{1}{\sqrt{2 \pi} \sigma} \e^{-\frac{\l( X{-}\mu \r)^2}{2\sigma^2}} \l( 1 + \erf \l( \frac{\alpha \l( X-\mu \r)}{\sqrt{2} \sigma} \r) \r),
\end{equation} 
where $\mu$ is a location parameter related to the mean of the distribution, $\sigma$ characterizes the width of the distribution, $\alpha$ controls the skewness of the distribution, and
\begin{equation}
\erf \l( z \r) \equiv \frac{2}{\sqrt{\pi}} \int_0^z dt\, \e^{-t^2}.
\end{equation}  
%
\begin{figure}[b]
\centering
\includegraphics[width=\linewidth]{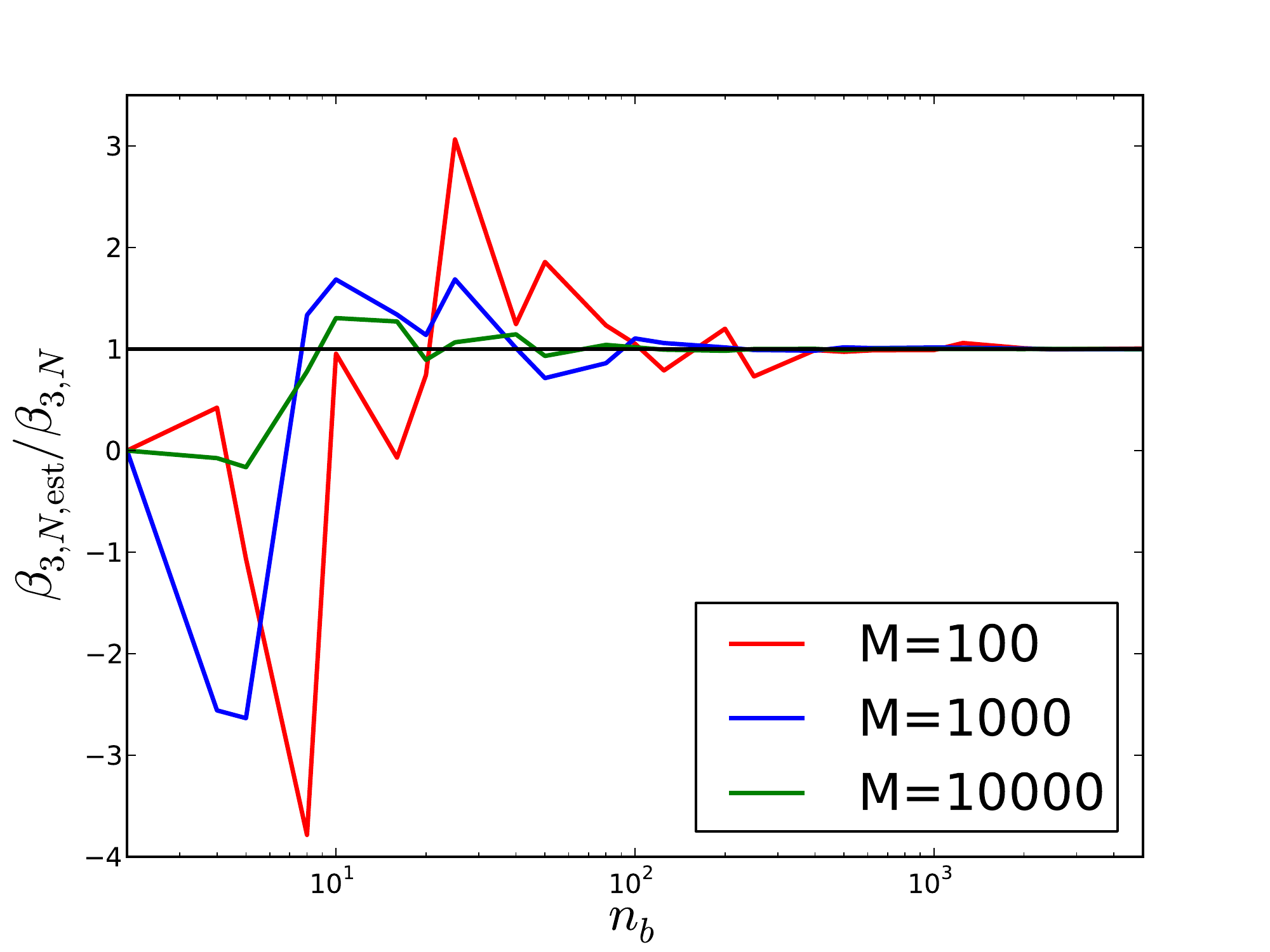} 
\caption{The estimated skewness $\beta_{3,N,\mathrm{est}}$, compared with the exact skewness $\beta_{3,N}$ of the sampled distribution, as a function of $n_b$ and M.  \label{Fig5:beta3_vs_nb_est}}
\end{figure}%
%
For $\alpha=0$, $\mu$ and $\sigma$ correspond to the mean and standard deviation of $P$, respectively.  In order to illustrate the generality of our result, our choice of parameters is somewhat arbitrary and entirely without physical motivation: we take $\mu=17$, $\sigma=3$, and $\alpha=10$.  The properties of this distribution are, for our choice of parameters:
\begin{eqnarray}
\mu_{\mathrm{true}} &\equiv& \int^{\infty}_{-\infty} dx\, x P(\mu,\sigma,\alpha;x) \nonumber\\
	&=& \mu + \alpha \sigma \sqrt{\frac{2}{\pi(1+\alpha^2)}} \approx 19.3818 \\
\sigma^2_{\mathrm{true}} & \equiv & \int^{\infty}_{-\infty} dx\, \l(x - \mu \r)^2 P(\mu,\sigma,\alpha;x) \nonumber\\
	&=& \sigma^2 \l( 1-\frac{2\alpha^2}{\pi \l( 1+\alpha^2 \r)} \r) \approx 3.32715 \\
\beta_{3,\mathrm{true}} &\equiv & \frac{M_{3,{\mathrm{true}}}}{\sigma^3_{\mathrm{true}}} \nonumber\\
	&=& \frac{\sqrt{2} (4-\pi) \alpha^3}{\l( \pi + (\pi-2)\alpha^2 \r)^{3/2}} \approx 0.955557 \\
\beta_{4,\mathrm{true}}-3 & \equiv & \frac{M_{4,{\mathrm{true}}}}{\sigma^4_{\mathrm{true}}}-3 \nonumber\\
	&=& \frac{8(\pi-3)\alpha^4}{\l( \pi + (\pi-2)\alpha^2 \r)^2} \approx 0.823244
\end{eqnarray}
The corresponding statistics for our specific measured sample of $N=10,\!000$ events
were
\begin{eqnarray}
\avg{X}_N & \equiv & 19.3588 
\label{exact_mu}\\
\sigma^2_N & \equiv & 3.22043 
\label{exact_sig2}\\
\beta_{3,N} & \equiv & 0.966647 
\label{exact_b3}\\
\beta_{4,N}-3 & \equiv & 1.00149
\label{exact_b4m3}
\end{eqnarray}
Using the methods defined by \eqref{defn_of_method}, \eqref{secondcentralmomentdefn} and \eqref{M3_est}-\eqref{beta4m3_defn} to estimate these sample statistics, for different values of $n_b$ and $M$, we can plot the convergence of these estimates to the exact values as functions of increasing $n_b$, for $M \in \l\lbrace 100,1000,10000 \r\rbrace$.  This is shown in Figs.~\ref{Fig4:sig2_vs_nb_est}--\ref{Fig6:beta4m3_vs_nb_est}.

\begin{figure}
\centering
\includegraphics[width=\linewidth]{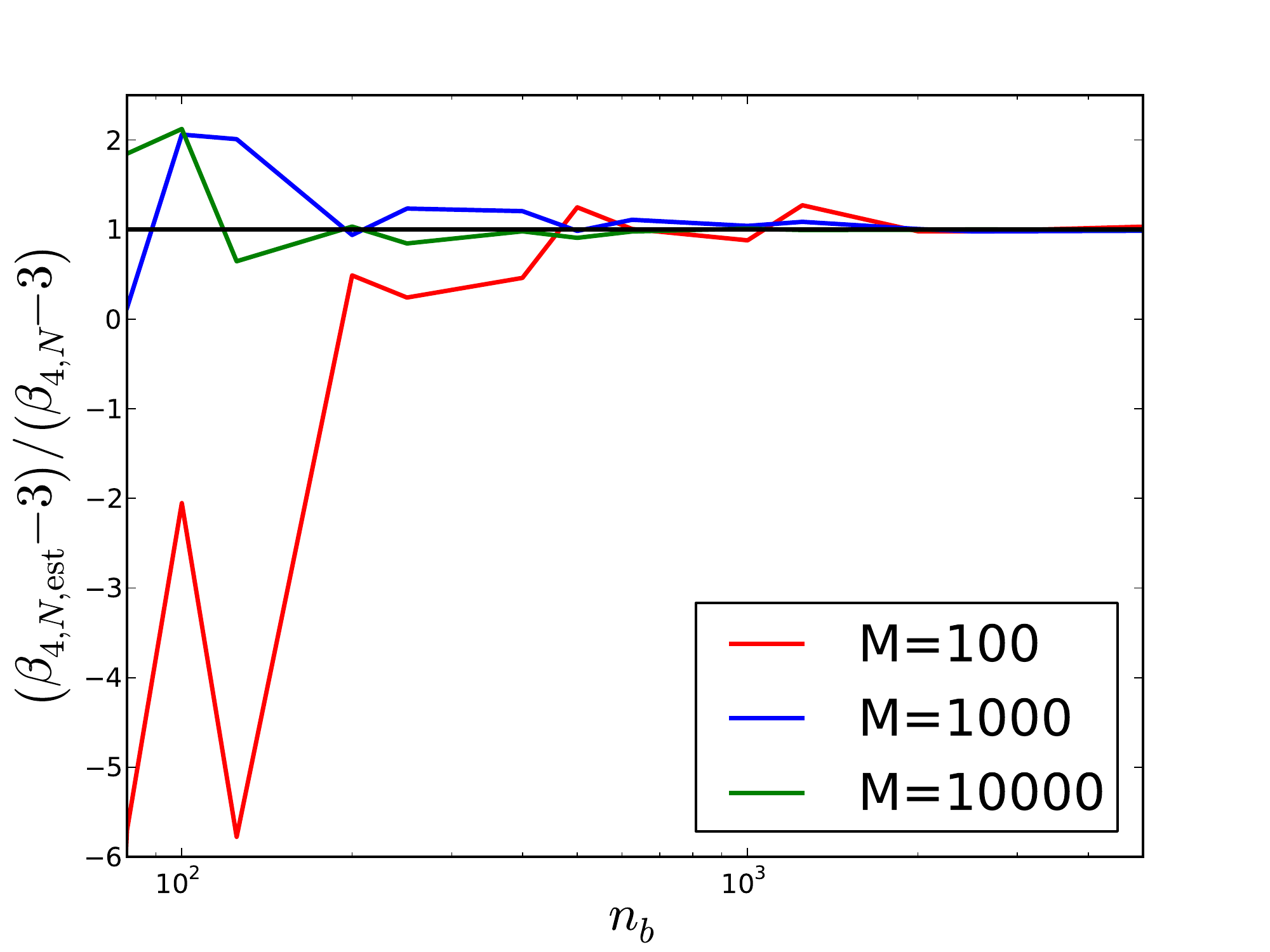} 
\caption{The estimated excess kurtosis $\beta_{4,N,\mathrm{est}}-3$, compared with the exact value $\beta_{4,N}-3$ of the sampled distribution, as a function of $n_b$ and M.  For aesthetic purposes, we have omitted the large fluctuations of the $M=100$ (red) curve for $n_b<80$. 
\label{Fig6:beta4m3_vs_nb_est}}
\end{figure}

All of three figures reveal a consistent trend: by increasing the number of bins $n_b$ and/or the number of binnings $M$, we can extract with good accuracy the variance, the skewness and the excess kurtosis of an event-by-event distribution in the observable $X$.  Improving the accuracy of this extraction comes with the numerical expense associated with increasing $n_b$ and/or $M$, exacerbated by the fact that for increasing $n_b$ one is decreasing the number of events-per-bin $n = N_{ev}/n_b$, resulting in larger finite-number statistical fluctuations of the bin averages. For a finite size $N$ of the total ensemble of measured events, there is therefore a maximal amount of information about the underlying $X$-distribution that can be extracted from the data, even in the limit of infinite computational resources.  Experimental HBT analyses typically are based on event samples of size $10^6-10^7$ rather than $10^4$. The first three moments of the HBT radii distributions should thus be accessible with statistical precision of better than 1\%, which is significantly below the typical systematic uncertainties associated with HBT measurements that are unrelated to the method discussed here. At this level of precision, these three moments may already be able to provide insights into the nature of the physics lying at the origin of the HBT radii fluctuations and help to further constrain the material properties of the quark-gluon plasma created in relativistic heavy-ion collisions \cite{Plumberg:2015eia}.

\subsection{Realistic application}
\label{sec8c}

We finally show how the combination of methods demonstrated in the previous two subsections can be used to obtain physically interesting results. For the same 5,000 hydrodynamic events considered in Sec.~\ref{sec4}, we use an azimuthally averaged, ensemble-averaged correlation function method to estimate the \textit{coefficient of variation} (or \textit{relative variance}) $\sigma_{ij}/R^2_{ij}$ of the event-by-event radii.  The potential theoretical significance of the relative variance in heavy-ion collisions has been recently explored in \cite{Plumberg:2015eia}.  Since the HBT radii are known theoretically on an event-by-event basis, we can compute this quantity exactly from Eqs.~\eqref{defn_of_true_mean} and \eqref{defn_of_true_variance}.  We now demonstrate how these exact quantities may be reliably estimated by the combination of mean estimation and variance estimation that we have presented above.

\begin{figure}
\includegraphics[width=\linewidth]{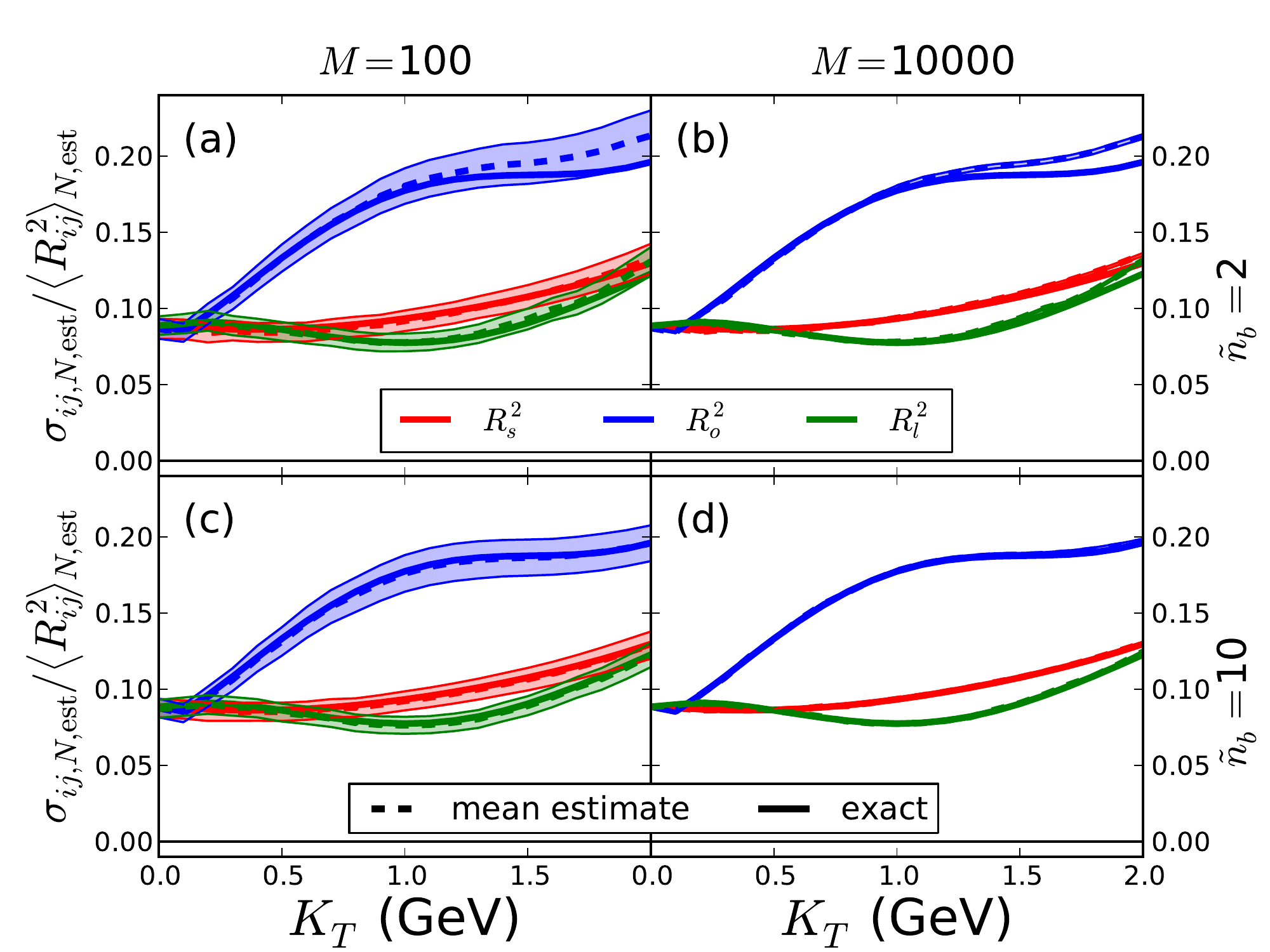}
\caption{For $N_{\ev}=5\!000$ and $n_b=2$, the left hand panels show the estimates for the relative variance for $R^2_s$ (red), $R^2_o$ (blue), and $R^2_l$ (green) for $M=100$, while the right hand panels show the same results for $M=10,\!000$.  Similarly, the upper row shows our estimated results for only $\tilde{n}_b = 2$, while the bottom row shows the same results for $\tilde{n}_b = 10$. The solid lines are the exact results for the ensemble of 5,000 events, while the dashed lines represent the average estimate we obtain after 100 random iterations of our method, and the shaded bands indicate the resulting standard deviation of our estimate.
\label{Fig7:finalresults}
}
\end{figure}

In detail, the combined algorithm works in the following way.  We begin with the same ensemble of $N_{\ev}=5,\!000$ events that was already used in Sec.~\ref{sec4}, and go through $M$ iterations of binning them into $n_b$ bins.  For each bin, we estimate the \textit{arithmetic} bin average by \textit{sub}dividing each bin into $\tilde{n}_b$ sub-bins, and using the method presented in Sec.~\ref{sec5}.  Once the arithmetic average for each of the $n_b$ bins is known, we use the method presented in Sec.~\ref{sec6} to estimate the variance $\sigma^2_{ij}$.

We show the results of this combined procedure in Fig.~\ref{Fig7:finalresults}.  Since each of the $M$ binning iterations requires a random partition of $N_{\ev}$ events into $n_b$ bins, our estimates will tend to exhibit some variability, particularly if $M$ is small (say, of order 100).  To quantify the resulting uncertainty of the estimates shown in Fig.~\ref{Fig7:finalresults}, we compute the estimates 100 different times, and then compute the mean and variance of these estimates.  The mean estimates are shown as dashed lines in Fig.~\ref{Fig7:finalresults}, and the shaded bands represent the $\pm1\sigma$ variability of our estimate.

For $\tilde{n}_b = 2$, the mean estimate clearly disagrees with the exact result for $K_T \geq 1$ GeV. This bias is reduced by increasing the number of sub-bins: taking $\tilde{n}_b=10$ reduces the bias of our estimation procedure almost to zero.  Similarly, with $M=100$, the variability of our estimation procedure is noticeable, but for $M=10,\!000$, the widths of the shaded bands are almost negligible. Thus, we see that increasing the number of sub-bins $\tilde{n}_b$ (top row vs. bottom row) results in the ability to decrease the bias in our estimate of the exact result, while increasing the total number of binning iterations $M$ (left column vs. right column) reduces the overall variability of our estimate.  For $n_b = 2$, $\tilde{n}_b = 10$ and $M=10,\!000$, the statistical uncertainty of our estimation procedure effectively disappears, and the methods presented in this paper provide us with a reliable way of accessing statistical moments of the ensemble distribution of the HBT radii.

\section{Outlook}
\label{sec9}

In this paper, we presented a method for extracting from experimental data properties of the event-by-event distributions of HBT radii that characterize heavy-ion collisions, by estimating their central moments. This method allows to overcome the current restriction of published HBT results to ensembled-averaged experimental data, and opens a way to experimentally access valuable information contained in the event-by-event fluctuations of HBT radii that complements analogous information from fluctuations of the momentum spectra and their associated anisotropic flow coefficients. The proposed method works for both azimuthally averaged and azimuthally sensitive HBT analyses. It requires large, but not exorbitant event statistics.

With this work, we promote the area of HBT interferometry of heavy-ion collisions to a new level that permits the systematic investigation of the statistical properties of event-by-event fluctuations of interferometric signatures and thereby advances this subfield to a similar level of sophistication as established over the last decade for momentum-space heavy-ion observables. We expect event-by-event HBT analyses to bear similarly rich fruit as has been recently harvested from studying event-by-event fluctuations of multiplicities and collective flow signatures.


\acknowledgments
C.P. would like to thank Jai Salzwedel and Michael Lisa for many valuable conversations which helped to clarify the ideas presented here, as well as for their continued interest in this research. This work was supported by the U.S. Department of Energy, Office of Science, Office of Nuclear Physics under Awards No. \rm{DE-SC0004286} and (within the framework of the JET Collaboration) \rm{DE-SC0004104}.

\begin{appendix}
\begin{widetext}
\section{Derivation of the variance estimator} 
\label{App:AppendixA} 

In this appendix, we prove that \eqref{defn_of_method} reduces to \eqref{defn_of_true_variance} when $M$ is taken to be the total number $M_{\mathrm{max}}(N,n_b)$ of possible distinct ways to sort $N$ events into $n_b$ bins of $n \equiv N/n_b$ events each.  To do this, we first expand the righthand side of \eqref{defn_of_true_variance}:
\begin{equation}
\sigma^2_{\mcO,N} = \frac{1}{N{-}1} \sum^{N}_{i=1}\left( \mcO^2_i - \avg{\mcO}_N^2 \right)
	= \frac{1}{N{-}1} \left( \left( \sum^{N}_{i=1} \mcO^2_i \right) - \frac{1}{N} \left( \sum^{N}_{i=1} \mcO_i \right)^2 \right)
	= \frac{1}{N} \sum^{N}_{i=1} \mcO^2_i - 2 \sum^{N}_{i=1}  \sum^{N}_{j=i+1} \frac{\mcO_i \mcO_j}{N(N{-}1)} \label{reworkedLHS}
\end{equation}  
We see that this expression consists of ``quadratic terms'' (the first sum) and ``cross terms'' (the second sum).  The maximal number of independent ways to distribute the total ensemble of $N$ events into $n_b$ bins of size $n=N/n_b$ is
\begin{eqnarray}
\Mmax(N,n_b) &\equiv & {N \choose n} {N{-}n \choose n} \cdots {2n \choose n} = \frac{N!}{(n!)^{n_b}} =\frac{N!}{\l(\l( N/n_b \r)!\r)^{n_b}}. \label{Mmaxdefn}
\end{eqnarray}  
For $M=\Mmax$, we may write the righthand side of \eqref{defn_of_method} as
\begin{equation}
\frac{N}{n_b (n_b{-}1)\Mmax} \sum^{\Mmax}_{j=1} \sum^{n_b}_{k=1} 
\left(\avg{\mcO}^2_{j,k}{-}\avg{\mcO}_N^2 \right) = \frac{N}{\Mmax (n_b{-}1)} \sum^{\Mmax}_{j=1} \left(\avg{\mcO}^2_{j} - \avg{\mcO}_N^2 \right), 
\label{reworkedRHS}
\end{equation} 
where
\begin{equation}
\sum^{\Mmax}_{j=1} \avg{\mcO}^2_{j,1} = \sum^{\Mmax}_{j=1} \avg{\mcO}^2_{j,2} = \cdots = \sum^{\Mmax}_{j=1} \avg{\mcO}^2_{j,n_b} \equiv \sum^{\Mmax}_{j=1} \avg{\mcO}^2_{j}.  \label{app_simplification}
\end{equation}  
This equality holds by definition when $\Mmax$ exhausts all binning possibilities, because the different sums in \eqref{app_simplification} only differ by permutations of their summands.

Introducing the notation $\mcO^{(j)}_k$ to represent the $k$th event observable in the $j$th iteration,%
\footnote{%
	Of course, since each iteration can be thought of as a partition into $n_b$ bins 
	of a random permutation of all $N$ events in the ensemble, the $k$th event 
	observable in the $j$th iteration will generally not be the same as $k$th event
	observable of the $j+1$st iteration.}
we can expand both of the terms in \eqref{reworkedRHS} as follows, to bring them into the form \eqref{reworkedLHS}. Using \eqref{app_simplification}, the first term in \eqref{reworkedRHS} becomes
\begin{eqnarray}
\frac{N}{\Mmax(n_b{-}1)} \sum^{\Mmax}_{j=1} \avg{\mcO}^2_{j} &=& \frac{N}{\Mmax(n_b{-}1)} \sum^{\Mmax}_{j=1} \l( \frac{1}{n}\sum^n_{k=1} \mcO^{(j)}_k \r)^2 \nonumber\\
	&=& \frac{N}{n^2 \Mmax (n_b{-}1)} \sum^{\Mmax}_{j=1} \sum^n_{k=1} \sum^n_{k'=1} \mcO^{(j)}_k \mcO^{(j)}_{k'} \nonumber\\
	&=&  \frac{N}{n^2 \Mmax (n_b{-}1)} \sum^{\Mmax}_{j=1} \l[ \sum^n_{k=1} \l( \mcO^{(j)}_k \r)^2 + 2 \sum^n_{k=1} \sum^n_{k'=1} \mcO^{(j)}_k \mcO^{(j)}_{k'} \r] \nonumber\\
	& \equiv & \frac{N}{n^2 \Mmax (n_b{-}1)} \l( \alpha_1 \sum^{N}_{i=1} \mcO^2_k + 2 \alpha_2 \sum^{N}_{k=1} \sum^{N}_{k'=1} \mcO_k \mcO_{k'} \r)
\end{eqnarray} 
$\alpha_1$ is a degeneracy factor which counts the number of times any given event falls into the first bin; with our choice of $\Mmax$, this is just $\Mmax/n_b$.  Similarly, $\alpha_2$ is a degeneracy factor which counts the number of iterations for which two different event both fall into the first bin; it is given by 
\begin{equation}
\alpha_2 \equiv {N{-}2 \choose n{-}2} {N{-}n \choose n} \cdots {2n \choose n} = \frac{\Mmax n (n{-}1)}{N(N{-}1)}
\end{equation}
The second term in \eqref{reworkedRHS} is less subtle:
\begin{eqnarray}
\frac{N}{\Mmax(n_b - 1)} \sum^{\Mmax}_{j=1} \avg{\mcO}_N^2 &=& \frac{1}{N(n_b{-}1)} \sum^{N}_{i=1}\sum^{N}_{j=1} \mcO_i \mcO_j \nonumber\\
	&=& \frac{1}{N(n_b{-}1)} \sum^{N}_{i=1} \mcO^2_i + \frac{2}{N} \sum^{N}_{i=1}\sum^{N}_{j=i+1} \mcO_i \mcO_j
\end{eqnarray}
Combining these results, we obtain 
\begin{eqnarray}
\frac{N}{M(n_b{-}1)} \sum^{\Mmax}_{j=1} \left(\avg{\mcO}^2_{j}{-}\avg{\mcO}_N^2 \right) &=& \l( \frac{N{-}n}{N n(n_b{-}1)} \r) \sum^{N}_{i=1} \mcO^2_i + \frac{2}{n_b{-}1} \left( \frac{n{-}1}{n(N{-}1)} - \frac{1}{N} \right) \sum^{N}_{i=1}  \sum^{N}_{j=i+1} \mcO_i \mcO_j \nonumber\\
&-& \frac{1}{N} \sum^{N}_{i=1} \mcO^2_i - \frac{2}{N} \sum^{N}_{i=1}\sum^{N}_{j=i+1} \mcO_i \mcO_j \nonumber\\
	&=& \frac{1}{N} \sum^{N}_{i=1} \mcO^2_i - \frac{2}{N(N{-}1)} \sum^{N}_{i=1}  \sum^{N}_{j=i+1} \mcO_i \mcO_j,
\end{eqnarray} 
which is identical to \eqref{reworkedLHS}. This establishes that $\sigma^2_{\mcO,N,\mathrm{est}}$ from \eqref{defn_of_method} reduces to $\sigma^2_{\mcO,N}$ from \eqref{defn_of_true_variance} when $M=\Mmax$.
\end{widetext}
\section{Two derivations of $R^2_{\avg{ij}}$} 
\label{App:AppendixB}

In this appendix, we present two different ways of understanding the relationship between the event-by-event HBT radii and the experimentally measured, ensemble-averaged radii, $R^2_{\avg{ij}}$.  First, we show how the relationship \eqref{true_R2ij_from_corrfunc} arises from the combination of \eqref{corrfuncSEdefn} and \eqref{corrfuncENSAVG0defn}.  We then show that precisely the same result holds for the radii extracted from \eqref{Cavg_corrfunc_def} via the source-variances method, motivating us to treat \eqref{Cavg_corrfunc_def} as the most physically accurate way of accounting for the contribution of event-by-event fluctuations to the experimental correlator \eqref{true_R2ij_from_corrfunc}.

We begin by introducing the convenient shorthand
\begin{equation}
N^{(k)}_{1,2} \equiv E_1 E_2 \frac{d^6N^{(k)}}{d^3 p_1 d^3 p_2}
\end{equation} 
and
\begin{equation}
N^{(k)}_{\ell} \equiv E_{\ell} \frac{d^3N^{(k)}}{d^3p_\ell},\quad {\ell}=1,2,
\end{equation}
where $k$ labels a particular event in the ensemble.  In this notation, the correlation function \eqref{corrfuncSEdefn} for the $k$th event is written
\begin{equation}
C^{(k)} \equiv \frac{N^{(k)}_{1,2}}{N^{(k)}_1 N^{(k)}_2} \equiv 1 + \exp \l(-\sum_{i,j = o, s, l} R^2_{ij} q_i q_j \r), \label{SEcorrs}
\end{equation} 
where we have suppressed the functional momentum dependence for the sake of clarity.  On the other hand, the correlator defined from the ensemble-averaged one-particle and two-particle spectra is written
\begin{eqnarray}
\avg{C}_{\ev} &=& \frac{\avg{N_{1,2}}}{\avg{N_1}\avg{N_2}} \nonumber\\
	&=& \frac{\frac{1}{N_{\ev}}\sum_{k=1}^{N_{\ev}}N^{(k)}_{1,2}}{\l( \frac{1}{N_{\ev}}\sum_{k=1}^{N_{\ev}} N^{(k)}_1 \r)\l( \frac{1}{N_{\ev}}\sum_{k=1}^{N_{\ev}} N^{(k)}_2 \r)} \nonumber\\
	&=& \frac{\frac{1}{N_{\ev}}\sum_{k=1}^{N_{\ev}} C^{(k)} N^{(k)}_1 N^{(k)}_2}{\l( \frac{1}{N_{\ev}}\sum_{k=1}^{N_{\ev}} N^{(k)}_1 \r)\l( \frac{1}{N_{\ev}}\sum_{k=1}^{N_{\ev}} N^{(k)}_2 \r)} \nonumber\\
	& \equiv & \frac{1}{N_{\ev}}\sum_{k=1}^{N_{\ev}} w_k C^{(k)}, \label{EAcorr_sum_SEcorrs}
\end{eqnarray} 
where we defined
\begin{equation}
w_k \equiv \frac{N^2_{\ev} N^{(k)}_1 N^{(k)}_2}{\l( \sum_{k=1}^{N_{\ev}} N^{(k)}_1 \r)\l( \sum_{k=1}^{N_{\ev}} N^{(k)}_2 \r)} = \frac{N^{(k)}_1 N^{(k)}_2}{\bar{N}^2}.
\end{equation} 
The last step follows in the smoothness approximation \cite{Pratt:1997pw} if we additionally define $\bar{N} \equiv \avg{N_1} = \avg{N_2}$.

Although the single event correlators \eqref{SEcorrs} reflect the traditionally imposed normalizations
\begin{equation}
\lim_{\vec{q} \rightarrow 0} C^{(i)}(\vec{q},\vec{K}) = 2,\quad \lim_{\vec{q} \rightarrow \infty} C^{(i)}(\vec{q},\vec{K}) = 1, \label{SEcorr_lims}
\end{equation} 
the ensemble-averaged correlator defined by Eq.~\eqref{EAcorr_sum_SEcorrs} leads to a different overall normalization:
\begin{equation}
\lim_{\vec{q} \rightarrow 0} \avg{C}_{\ev}(\vec{q},\vec{K}) = 2\frac{\avg{N^2}}{\bar{N}^2},\quad \lim_{\vec{q} \rightarrow \infty} \avg{C}_{\ev}(\vec{q},\vec{K}) = \frac{\avg{N^2}}{\bar{N}^2}.  \label{EAcorr_lims}
\end{equation}  
Consequently, we fit \eqref{EAcorr_sum_SEcorrs} to a Gaussian form with an additional normalization factor:
\begin{eqnarray}
\avg{C}_{\ev} \equiv C_0 \l( 1 + \exp \l(-\sum_{i,j = o, s, l} R^2_{\avg{ij}} q_i q_j \r) \r), \label{EAcorr_normd_functional}
\end{eqnarray} 
where we have to set $C_0 \equiv \avg{N^2}/\bar{N}^2$ in order to satisfy \eqref{EAcorr_lims}.  Then, equating \eqref{EAcorr_normd_functional} with \eqref{EAcorr_sum_SEcorrs} and requiring their curvatures (with respect to $q$) to be the same as $\vec{q} \rightarrow 0$ leads immediately to the expression
\begin{equation}
\frac{\avg{N^2}}{\bar{N}^2}R^2_{\avg{ij}} = \frac{\avg{N^2 R^2_{ij}}}{\bar{N}^2},
\end{equation} 
which is simply Eq.~\eqref{true_R2ij_from_corrfunc}.

We can derive this same result by a slightly different treatment.  Instead of directly requiring the ensemble-averaged correlator to satisfy certain normalization constraints, we now show how to obtain the source-variance HBT radii from the physical correlation function \eqref{Cavg_corrfunc_def} which correctly incorporates the complete effects of event-by-event fluctuations.  By utilizing the source-variance method, we choose to approximate the correlation function as a Gaussian in $\vec{q}$, which means the corresponding radii may be extracted as above by simply computing the curvature of the correlator as $\vec{q} \rightarrow 0$:
\begin{equation}
R^2_{\avg{ij}} \equiv -\frac{1}{2\l( \avg{C}_{\ev} - 1 \r)} \left.\frac{\partial ^2 \avg{C}_{\ev}}{\partial q_i \partial q_j} \right| _{\vec{q}\rightarrow 0}. \label{curvature_R2ij_defn}
\end{equation} 
Here, the factor of $\avg{C}_{\ev} - 1$ in the denominator of this expression account for the eventwise multiplicity fluctuations of $N$.  To see this, consider an ensemble of emission functions which differ only in overall normalization $N$:
\begin{equation}
S_N(x,K) \equiv N S_0(x,K),
\end{equation} 
where $S_0(x,K)$ has the same Gaussian form for every event in the ensemble.  If the fluctuations of normalization are governed by a probability distribution $P(N)$, then we can write the ensemble-averaged emission function $\bar{S}(x,K)$ as
\begin{eqnarray}
\bar{S}(x,K) &\equiv & \avg{S_N(x,K)}_{\ev} \nonumber\\
	&=& S_0(x,K) \l( \int dN \, P(N)\, N \r) \nonumber\\
	&=& \bar{N} S_0(x,K),
\end{eqnarray} 
where $\bar{N}$ is the average normalization for the ensemble. Then, using Eqs.~\eqref{Cbar_corrfunc_def} and \eqref{Cavg_corrfunc_def}, it is straightforward to show that
\begin{equation}
\bar{C}(q,K) = 1+ \frac{\l|\int d^4x\,\e^{i q \cdot x}S_0(x,K)\r|^2}{\l|\int d^4x\, S_0(x,K)\r|^2}
\end{equation} 
and
\begin{eqnarray}
\avg{C}_{\ev}(\vec{q}, \vec{K}) &\equiv & 1 + \frac{ \l< \l| \int d^4 x \, \e^{i q \cdot x} S_N(x,K) \r|^2 \r>_{\ev}}{\l| \int d^4 x \, \bar{S}(x,K) \r|^2} \nonumber\\
	&=& 1+\frac{\avg{N^2}_{\ev}}{\bar{N}^2 }\frac{\l|\int d^4x\,\e^{i q \cdot x}S_0(x,K)\r|^2}{\l|\int d^4x\, S_0(x,K)\r|^2} \nonumber\\
	&=& 1+ \frac{\avg{N^2}_{\ev}}{\bar{N}^2 } \l( \bar{C}{-}1 \r)
\end{eqnarray}  
This result demonstrates that, although $\avg{C}_{\ev}$ and $\bar{C}$ both have the same dependence on $q$ and therefore the same HBT radii, $\bar{R}^2_{ij} = R^2_{\avg{ij}}$, the curvatures of the two correlators do not agree as $\vec{q}\rightarrow 0$: rather, they differ by an additional factor of $\avg{N^2}_{\ev}/\bar{N}^2$, which is equal to $\avg{C}_{\ev}\l(\vec{q} \rightarrow 0\r)-1$.  Consequently, we define the radii in terms of the appropriately normalized curvature of the correlator at the origin.%
\footnote{%
	Strictly speaking, we should apply the same correction factor to the first 
	correlator $\bar{C}$.  However, this would prove unnecessary, as it follows 
	by definition from Eq.~\eqref{Cbar_corrfunc_def} that 
	$\bar{C}\l( \vec{q} \rightarrow 0 \r)-1 = 1$.}
Using Eq.~\eqref{curvature_R2ij_defn}, we can then write
\begin{eqnarray}
R^2_{\avg{ij}}
	&=&  -\frac{1}{2\avg{N^2}_{\ev} N_{ev}} \sum^{N_{ev}}_{k=1} \int d^4 x d^4 x' \nonumber\\
	& & \left. \quad \quad \times \frac{\partial ^2 }{\partial q_i \partial q_j} \e^{i q \cdot (x-x')} S_k(x,K) S_k(x',K) \right| _{\vec{q}\rightarrow 0}
\nonumber\\
	&=& \frac{1}{2\avg{N^2}_{\ev} N_{ev}} \sum^{N_{ev}}_{k=1} \int d^4 x d^4 x' S_k(x,K) S_k(x',K) \nonumber\\
    & & \times \big[ \left( x_i{-}\beta_i t \right) - \left( x'_i{-}\beta_i t' \right) \big]
        \big[ \left( x_j{-}\beta_j t \right) - \left( x'_j{-}\beta_j t' \right) \big]
\nonumber\\
	&=& \frac{1}{N_{ev}} \sum^{N_{ev}}_{k=1} \frac{N^2_k}{\avg{N^2}_{\ev}} 
        \l( \avg{ \left( x_i{-}\beta_i t \right)\left( x_j{-}\beta_j t \right) } _{S_k} \r. \nonumber\\
    & & \l. \quad \quad \quad \quad \quad \quad - \avg{ \left( x_i{-}\beta_i t \right)} _{S_k} \avg{ \left( x_j{-}\beta_j t \right) } _{S_k} \r)
\nonumber\\
	&=& \frac{\avg{N^2 R^2_{ij}}_{\ev}}{\avg{N^2}_{\ev}},
\end{eqnarray} 
where, in the last step, we have simplified our result by means of Eq.~\eqref{svHBT_defn}.  This result is, again, seen to be equivalent to \eqref{true_R2ij_from_corrfunc}.

This means that a consistent extraction of the ensemble-averaged HBT radii should account for the fact that more pairs of identical particles will come from events with larger total charged multiplicities $dN^{\mathrm{ch}}/d\eta$, leading to the HBT radii for \textit{these} events being more heavily represented in the final, ensemble-averaged radii.  Consequently, the true average radii are in fact a \textit{weighted} average over the event-by-event HBT radii, where the weighting factor of $N^2/\avg{N^2}$ represents the fraction of all pion pairs used in the construction of \eqref{corrfuncENSAVG0defn} that come from an event with overall normalization $N$.

\end{appendix}



\end{document}